\documentclass[a4,10pt]{article}

\usepackage{amsmath,amsthm,amssymb,bm}
\usepackage[dvipdfmx]{graphicx}
\usepackage[dvipdfmx]{color}
\usepackage{cite}

\allowdisplaybreaks[3]

\makeatletter
  
  \@addtoreset{equation}{section}
\makeatother

\def\be{\begin{equation}}
\def\ee{\end{equation}}
\def\ba{\begin{eqnarray}}
\def\ea{\end{eqnarray}}

\newcommand{\hK}{\tilde K}

\def\smu{\smash{\mu}}
\def\tg{{\tilde g}}

\newcommand{\hook}
{\raisebox{-0.35ex}{\makebox[0.6em][r]{\scriptsize $-$}}
\hspace{-0.15em}\raisebox{0.25ex}{\makebox[0.4em][l]{\tiny $|$}}}

\begin{document}

\title{On symmetry operators for the Maxwell equation 
on the Kerr-NUT-(A)dS spacetime}

\author{Tsuyoshi Houri$^1$, Norihiro Tanahashi$^2$, Yukinori Yasui$^3$\\[0.5cm]
$^1$National Institute of Technology, Maizuru College, Kyoto 625-8511, Japan,\\
$^2$Institute of Mathematics for Industry, Kyushu University, Fukuoka 819-0395, Japan,\\
$^3$Institute for Fundamental Sciences, Setsunan University, Osaka 572-8508, Japan}


\date{\today}

\maketitle

\begin{abstract}
We focus on the method recently proposed 
by Lunin and Frolov-Krtou\v{s}-Kubiz\v{n}\'{a}k
to solve the Maxwell equation on the Kerr-NUT-(A)dS spacetime 
by separation of variables. 
In their method, it is crucial that the background spacetime 
has hidden symmetries 
because they generate commuting symmetry operators 
with which the separation of variables can be achieved.
In this work we reproduce these commuting symmetry operators 
in a covariant fashion.
We first review the procedure known as the Eisenhart-Duval lift
to construct commuting symmetry operators for given equations of motion.
Then we apply this procedure to 
the Lunin-Frolov-Krtou\v{s}-Kubiz\v{n}\'{a}k (LFKK) equation.
It is shown that the commuting symmetry operators 
obtained for the LFKK equation
coincide with the ones previously obtained 
by Frolov-Krtou\v{s}-Kubiz\v{n}\'{a}k,
up to first-order symmetry operators 
corresponding to Killing vector fields.
We also address the Teukolsky equation
on the Kerr-NUT-(A)dS spacetime and 
its symmetry operator is constructed.
\end{abstract}

\newpage
\tableofcontents

\section{Introduction}
\label{sec:intro}

Black hole spacetimes and their perturbations have been a central topic in gravitational physics.
Typically the equations of motion for the perturbations become complicated.
It is crucial to use various techniques to simplify them.
The importance of such techniques become higher especially in, e.g., analysis on astrophysical processes around black holes.
Among such techniques, separation of variables in perturbation equations provides a way to simplify the problem 
because it reduces the perturbation equations given as partial differential equations 
into a set of ordinary differential equations parameterized by separation constants.

In this paper, we focus on the electromagnetic perturbation on the background
the rotating black hole spacetimes \cite{Kerr:1963ud,Myers:1986un},
which is governed by the Maxwell field equation 
\begin{equation}
 \nabla^a {\cal F}_{ab} = 0 \,.
\label{Maxwellequation}
\end{equation}
This equation becomes complicated due to the rotation of the background spacetime, 
and in general it cannot be solved without resorting to elaborate techniques.
For the Kerr spacetime in four dimensions,
this problem was resolved by Teukolsky \cite{PhysRevLett.29.1114,Teukolsky:1973ha} 
by expressing the equation based on the Newman-Penrose formalism and separating variables completely.

Generalizations of this technique to higher dimensions had been sought for since then, 
and rather recently a new method was proposed by Lunin \cite{Lunin:2017drx}, 
with which the mathematical structure behind the Teukolsky equation can be simplified 
and as a result separation of variables is accomplished for all modes.

In this technique, the hidden symmetry of the background spacetime was a key to achieve the separation of variables. 
Kerr spacetime and its higher-dimensional counterparts are associated with Killing tensors $K_{ab}$ satisfying
\begin{equation}
\nabla_{(a}K_{bc)}=0 \,,
\end{equation}
and they yield constants of motion preserved on geodesics once contracted with momentum $p^a$.
They are independent of those associated to explicit spacetime symmetry and Killing vectors, 
and in this sense the Killing tensors correspond to hidden symmetry of the background spacetime.
In Lunin's work, this tensor was used to introduce special coordinates that helped separating the variables of the perturbation equations.

About this technique, Frolov, Krtou\v{s} and Kubiz\v{n}\'{a}k \cite{Frolov:2018pys,Krtous:2018bvk} 
clarified 
the significance of the hidden symmetry of the spacetimes with regard to the separation of variables in the Maxwell equation.
the principal tensor
$h_{ab}$ defined as an anti-symmetric tensor satisfying
\begin{equation}
\nabla_c h_{ab} =
g_{ca} \xi_b - g_{cb} \xi_a ~,
\qquad
\xi_a = \frac{1}{D-1} \nabla^b h_{ba} ~ .
\label{principaltensor}
\end{equation}
It is known that the (off-shell) Kerr-NUT-(A)dS spacetime in arbitrary dimensions is the most general spacetime 
admitting this principal tensor, 
and the Kerr spacetime in four dimensions and Myers-Perry spacetime in higher dimensions are included in this family.

What they showed was that, by  expressing 
the vector potential ${\cal A}^a$ 
expressed with a scalar $\Phi$ and a constant $\beta$ as
\begin{equation}
{\cal A}^a = B^{ab}\nabla_b \Phi \,, 
\label{Aansatz}
\end{equation}
where $B^{ab}$ is the polarization tensor defined by
\begin{equation}
\left(g_{ab} - \beta h_{ab}\right)B^{bc} = \delta_a^c \,,
\label{B}
\end{equation}
the Maxwell equation in the Lorenz gauge
\begin{equation}
 \nabla_a {\cal A}^a = \nabla_a(B^{ab}\nabla_b \Phi) = 0
 \label{Lorenzgauge}
\end{equation}
can be reduced to a scalar-type equation for $\Phi$ given by
\begin{equation}
 \square \Phi + 2\beta \xi_a B^{ab}\nabla_b \Phi = 0 \,,
\label{LFKKeq}
\end{equation}
which we call the Lunin-Frolov-Krtou\v{s}-Kubiz\v{n}\'{a}k (LFKK) equation in this work.

They further showed that Eq.~(\ref{LFKKeq}) and also the Lorenz gauge condition~(\ref{Lorenzgauge}) can be expressed in terms of mutually commuting symmetry operators, and the separation of variables is established thanks to the commutativity of them.
This bland-new technique initiated active discussions on this subject (see e.g.\ \cite{Frolov:2018ezx,Dolan:2018dqv,Frolov:2018eza,Dolan:2019hcw,
Araneda:2019uwy,Lunin:2019pwz,Cayuso:2019vyh}).

In our work, we give a covariant description to the above-mentioned techniques, particularly to origin of the commuting symmetry operators.
For this purpose we employ
the Eisenhart-Duval lift \cite{10.2307/1968307,10.2307/1968433,PhysRevD.31.1841,Duval:1990hj,doi:10.1063/1.1899986}
(see e.g.\ \cite{Cariglia:2014ysa} for a review),
in which the original spacetime $(M,g_{ab})$ is supplemented with two additional dimensions,
and the differential operator $\square  + 2\beta \xi_a B^{ab}\nabla_b $ 
in (\ref{LFKKeq}) can be expressed simply as the d'Alembertian 
$\tilde \square = \tilde g^{AB} \tilde \nabla_A \tilde \nabla_B$
on the uplifted spacetime $(\tilde{M},\tilde{g}_{AB})$ in two higher dimensions.
Hence, with this technique we can reduce the LFKK equation 
to the massless Klein-Gordon equation by absorbing the additional term $2\beta \xi_c B^{cd}\nabla_d $
into the part of the higher-dimensional spacetime.

It turns out that the uplifted metric associated with the LFKK equation
falls into the standard form of the metrics 
whose geodesics are completely integrable \cite{MR1156532},
and correspondingly it admits  Killing vectors $\tilde L^{(i)A}$ and tensors $\tilde K_{AB}^{(i)}$ 
as many as the number of the spacetime dimensions.
It can be shown that the differential operators given by
$\tilde {\cal L}^{(i)}\equiv \tilde L^A \tilde \nabla_A$ and 
$\tilde {\cal K}^{(i)}\equiv \tilde \nabla_A \tilde K^{(i)AB} \tilde\nabla_B$
commute with the d'Alembertian  
$\bigl[\tilde \square,\tilde{\cal L}^{(i)}\bigr]
= 0 =\bigl[\tilde \square,\tilde{\cal K}^{(i)}\bigr]$
and also commute mutually as
$\bigl[\tilde{\cal L}^{(i)},\tilde{\cal L}^{(j)}\bigr] = 0$,
$\bigl[\tilde{\cal L}^{(i)},\tilde{\cal K}^{(j)}\bigr] = 0$,
and $\bigl[\tilde{\cal K}^{(i)},\tilde{\cal K}^{(j)}\bigr] = 0$.
Hence the differential operators $\tilde {\cal K}^{(i)}$
act as commuting symmetry operators of the massless Klein-Gordon equation upstairs
and provide commuting symmetry operators ${\cal L}^{(i)}$ and ${\cal K}^{(i)}$ 
of the LFKK equation downstairs.
Up to differences proportional to ${\cal L}^{(i)}$,
these operators coincide with those of \cite{Krtous:2018bvk}, in which only coordinate expressions of those operators were given.
Hence we devised a method to express the symmetry operators of \cite{Krtous:2018bvk} in a covariant manner and this is the main result of this work.

This paper is organized as follows.
We first propose a procedure to construct commuting symmetry operators for the general second-order differential equation of motion in Sec.~\ref{sec:sec2}.
We will take a geometric approach based on the Eisenhart-Duval lift to simplify the equation of motion, and 
then construct the Killing tensors associated to the uplifted metric to obtain the commuting symmetry operators.
As an illustration of our method, we apply the procedure to the Maxwell perturbations on the four-dimensional Kerr spacetime in Sec.~\ref{sec:sec3}.
Applying our procedure to the Teukolsky equation for the Maxwell field, the uplifted metric, the Killing tensors associated to it, and then the symmetry operators based on them are constructed in order.
In Sec.~\ref{sec:sec4}, we consider generalization to higher dimensions on the background of Kerr-NUT-(A)dS spacetime.
Briefly reviewing the properties of the background geometry and its hidden symmetry in sections~\ref{sec:KNAdS-Ddim} and \ref{sec:HiddenSymofKNUTAdS}, 
we apply our procedure to the LFKK equation on this background in Sec.~\ref{sec:lift-Ddim} to construct the commuting symmetry operators.
In Sec.~\ref{sec:comparison} we compare them with the commuting symmetry operators of \cite{Krtous:2018bvk} and find that they coincide with each other up to the lower-order differential operators associated to the Killing vectors. We also examine the Lorenz gauge condition in Sec.~\ref{sec:Lorenzgauge}.
Then we conclude this work with discussion in Sec.~\ref{sec:summary}.

Appendices are dedicated to technical details of this work.
The Killing equation of the base and uplifted spacetimes are summarized in appendix~\ref{app:Killingeq}, which becomes important to construct the symmetry operators from the Killing tensors.
Appendix~\ref{App:Teukolsky} is a review on the derivation of Teukolsky equation for Maxwell perturbations on the rotating black hole background based on the Hertz equation~\cite{Benn:1996su}.
In appendix~\ref{sec:anomaly-Ddim} 
we examine the commutativity condition for the LFKK equations, which must be satisfied for the symmetry operators constructed in Sec.~\ref{sec:sec4} to commute with each other.

\section{Construction of commuting symmetry operators}
\label{sec:sec2}

The Eisenhart-Duval lift
\cite{10.2307/1968307,10.2307/1968433,PhysRevD.31.1841,Duval:1990hj,doi:10.1063/1.1899986}
 is a technique of dealing with classical and quantum mechanical systems geometrically.
It is especially useful to find symmetries of equations of motion covariantly.
These symmetries are described by commuting symmetry operators
which map a solution to equations of motion into another one.
In this section, we review the procedure
to construct commuting symmetry operators for a given equation of motion
by using the Eisenhart-Duval lift.

\subsection{Eisenhart-Duval lift of a classical mechanical system}
\label{subsec:2.1}

We consider the motion of a charged particle 
with unit mass and charge $q$ on a $D$-dimensional 
Riemannian manifold $M$ with the metric ${\bm g} = g_{ab}dx^a dx^b$.
In the presence of vector and scalar potentials ${\bm A}=A_adx^a$ and $V$, 
the Hamiltonian is given by
\be
 H = \frac{1}{2}g^{ab}\left(p_a - qA_a\right)
 \left(p_b - qA_b\right) + V \,,
 \label{originalHamiltonian}
\ee
where $x^a$ are local coordinates on $M$ and
$p_a$ are momenta conjugate to $x^a$.
For such a system,
we consider the Eisenhart-Duval lift,
which is the uplift from $(M,{\bm g})$ with two directions
given by a timelike vector field $\partial/\partial u$
and a null vector field $\partial/\partial v$,
by introducing
the $(D+2)$-dimensional spacetime $\tilde{M} = M \times R^2$ with the metric
\begin{align}
\tilde{\bm{g}}
= \tg_{AB}d\tilde x^Ad\tilde x^B
= g_{ab}dx^a dx^b + 2 qA_a dx^a du + 2 du dv - 2 V du^2 \,,
\label{liftedmetric}
\end{align}
where $(\tilde x^A)=(x^a,u,v)$ are local coordinates on ${\tilde M}$.
Here, the small indices $a,b,\dots$ run over 1 to $D$,
while the capital indices $A,B,\dots$ run over 1 to $D+2$.
Particularly, $\tilde x^{D+1}=u$ and $\tilde x^{D+2}=v$.
We put tilde to quantities on $\tilde M$ besides the coordinates $\tilde x^A$
henceforth.
It should be noted that the components ${\tilde g}_{AB}$, i.e.,
\be
 \tg_{ab} = g_{ab}(x^c) \,, \quad
 \tg_{a u} = qA_a(x^c) \,, \quad
 \tg_{uv} = 1 \,, \quad
 \tg_{uu} = - 2V(x^c) \,,
\ee
are functions of $x^a$ only.
The Hamiltonian for geodesics
on the uplifted spacetime $(\tilde{M},\tilde{\bm{g}})$ is written as
\be
 \tilde{H} = \frac{1}{2}\tg^{AB}\tilde p_A\tilde p_B
 = \frac{1}{2}g^{ab} (p_a-qA_a p_v)(p_b -qA_b p_v)
+ V p_v^2 + p_u p_v \,,
\label{liftedHamiltonian}
\ee
where
\be
 \tg^{ab} = g^{ab} \,, \quad
 \tg^{a v}= -q g^{ab}A_b \,, \quad
 \tg^{uv} = 1 \,, \quad
 \tg^{vv} = q^2A_a A^a + 2V \,,
\ee
and we introduced the canonical momenta $(\tilde p_A) = (p_a, p_u, p_v)$ on $\tilde M$.
Since $g_{ab}$, $A_a$ and $V$ are functions of $x^a$ only,
$\partial/\partial u$ and $\partial/\partial v$ are Killing vector fields 
on $(\tilde{M},\tilde{\bm{g}})$;
hence, the corresponding momenta $p_u$ and $p_v$ are constants.
Putting $\tilde{H}=0$, $p_u=-E$ and $p_v=1$,
we recover the equations of motion for the Hamiltonian
 (\ref{originalHamiltonian}) with energy $H = E$
 from the Hamiltonian (\ref{liftedHamiltonian}). 
 Thus null geodesics on the uplifted spacetime ($\tilde{M},\tilde{\bm{g}}$) 
 project onto the solutions of the original system on $M$.

\subsection{Application to quantum mechanical systems}
\label{subsec:2.2}

In quantum mechanics, the equation of motion corresponding
to the system given by
the Hamiltonian (\ref{originalHamiltonian}) is obtained as
\begin{equation}
 {\cal H}\Phi \equiv \Big[g^{ab}(\nabla_a - iqA_a)
 (\nabla_b - iqA_b) -2 V\Big]\Phi = E \Phi \,,
 \label{originalSchEq}
\end{equation}
where $\nabla$ is the Levi-Civita connection on $(M,\bm{g})$.
In the Eisenhart-Duval lift,
the solutions to the equation of motion (\ref{originalSchEq})
can be reproduced from a particular class of
the solutions to the massless Klein-Gordon equation 
on the uplifted spacetime $(\tilde{M},\tilde{\bm{g}})$,
\be
 \tilde{\Box} \tilde \Phi = 0 \,,
 \label{liftedLaplaceEq}
\ee
where $\tilde{\Box} \equiv \tg^{AB}\tilde{\nabla}_A\tilde{\nabla}_B$ is 
the d'Alembertian with the Levi-Civita connection $\tilde{\nabla}$
on $(\tilde{M},\tilde{\bm{g}})$.
To see this, we shall calculate the d'Alembertian $\tilde{\Box}$ 
with the expression (\ref{liftedHamiltonian}), which leads to
\begin{align}
\tilde{\square}
&= \frac{1}{\sqrt{|\tg|}} \frac{\partial}{\partial \tilde x^A} \sqrt{|\tg|}\tg^{AB}
 \frac{\partial}{\partial \tilde x^B} \nonumber\\
&= \square
-2 qA^a \frac{\partial}{\partial x^a}\frac{\partial}{\partial v}
-q(\nabla_a A^a)\frac{\partial}{\partial v}
 +(q^2A_a A^a+2 V)\left( \frac{\partial}{\partial v} \right)^2
+2\frac{\partial}{\partial u}\frac{\partial}{\partial v} \,,
\end{align}
where $\Box \equiv g^{ab}\nabla_a\nabla_b$ is the Laplacian on $(M,\bm{g})$.
Restricting the solutions to Eq.\ (\ref{liftedLaplaceEq}) 
to the specific ones of the form
\be
 \tilde  \Phi = e^{iEu/2}e^{iv} \Phi({x^a}) \,, 
 \label{restrictedformofsolutions}
\ee
we have
\be
 \tilde{\Box} \tilde \Phi 
 = e^{iEu/2}e^{iv} \left({\cal H} -E\right) \Phi \,,
\ee
where ${\cal H}$ is the one defined by (\ref{originalSchEq}).
Hence, we find that, 
instead of solving the equation of motion (\ref{originalSchEq}),
we may solve the massless Klein-Gordon equation (\ref{liftedLaplaceEq}) 
on the uplifted spacetime $(\tilde{M},\tilde{\bm{g}})$.
This fact is the key in constructing symmetry operators of the equation of motion (\ref{originalSchEq}), 
as shown in the next subsections.

Below, we make several remarks on Eq.\ (\ref{originalSchEq}) 
which will be important in later discussion.

\begin{itemize}

\item
The Eisenhart-Duval lift may be used a little more flexibly.
Since discussion above is mostly unaffected by the choice of the signature of the metric $\bm{g}$, 
we may consider Eq.\ (\ref{originalSchEq}) as the equation on a $D$-dimensional pseudo-Riemannian manifold $(M,\bm{g})$.
Then the signature of
the uplifted $(D+2)$-dimensional metric $\tilde{\bm{g}}$ becomes
 ultrahyperbolic $(-,\dots,-,+,\dots,+)$, 
 since the signature of the metric corresponding to the additional dimensions is $(-,+)$.
In this paper, we apply the Eisenhart-Duval lift
to the Teukolsky equation in Sec.~\ref{sec:sec3}
and the LFKK equation in Sec.~\ref{sec:sec4}.
Both equations fit into the form (\ref{originalSchEq})
with the Kerr-NUT-(A)dS metric which is Lorentzian.

\item
Eq.\ (\ref{originalSchEq}) is written in the alternative form
\begin{equation}
 \Big(\Box -2 iq A^a \partial_a + F\Big)\Phi = E\Phi \,,
 \label{LFKK_old_form}
\end{equation}
where
\begin{eqnarray}\label{c2}
F = - iq \nabla^a A_a - q^2 A^a A_a - 2V.
\label{F_LFKK-Ddim}
\end{eqnarray}
This suggests that the Eisenhart-Duval lift
is applicable to a wide range of equations of motion.
For example, since the LFKK equation (\ref{LFKKeq}) is provided
 in the form (\ref{LFKK_old_form}),
we rewrite it into the form (\ref{originalSchEq})
in Sec.~4.

\item
Eq.\ (\ref{originalSchEq}) is covariant under the gauge transformation,
\begin{equation}
 A_a \to A_a' = A_a -iq^{-1} (d\Lambda)_a \,,
 \label{gauge_transf_SchEq}
\end{equation}
with an arbitrary function $\Lambda$
and $g_{ab}$ and $V$ unchanged.
Under this transformation, the differential operator ${\cal H}$ defined by
(\ref{originalSchEq}) is transformed as
\be
 {\cal H} \to
 {\cal H}' \equiv g^{ab}(\nabla_a - iqA'_a)
 (\nabla_b - iqA'_b) -2 V =
 e^\Lambda {\cal H} e^{-\Lambda} \,,
 \label{def_Hprime}
\ee
which shows that, if and only if $\Phi$ is a solution of
Eq.\ (\ref{originalSchEq}), $\Phi' \equiv e^\Lambda \Phi$
is a solution of the equation of motion 
\be
 {\cal H}'\Phi'=E\Phi' \,.
 \label{Hprimeeq}
\ee

\end{itemize}

\subsection{Separability and commuting symmetry operators}
\label{subsec:2.3}

The separability of Eq.\ (\ref{originalSchEq}) is intimately related 
to the existence of first- and second-order differential
operators ${\cal S}^{(i)}$
satisfying the commutation relations
\be
 \left[{\cal S}^{(i)},{\cal S}^{(j)}\right] = 0 \,,
 \qquad i,j=0,1,2,\dots
 \label{General_com_rel}
 \ee
where $[\,\, , \,]$ stands for the ordinary commutator $[A,B]\equiv AB-BA$ and ${\cal S}^{(0)}\equiv {\cal H}$.
The operators ${\cal S}^{(i)}$ that commute with ${\cal H}$, 
i.e., $[{\cal H},{\cal S}^{(i)}]=0$, are called symmetry operators
since they map a solution of Eq.\ (\ref{originalSchEq}) into another solution.
Equation~(\ref{General_com_rel}) also states that symmetry operators must commute with each other.

How are these symmetry operators related to the separability?
When Eq.\ (\ref{originalSchEq}) is solvable by separation of variables, 
the separation constants $s^{(i)}$ are related to 
the commuting symmetry operators ${\cal S}^{(i)}$ 
by the relation
\be
 {\cal S}^{(i)}\Phi = s^{(i)} \Phi \,.
 \label{rel-smo-sepconst}
\ee
Particularly, Eq.\ (\ref{rel-smo-sepconst}) for $i=0$ is equivalent to
Eq.\ (\ref{originalSchEq}) with $s^{(0)}=E$.
Later, we obtain the commuting symmetry operators
for the Teukolsky equation in Sec.~3 and the LFKK equation in Sec.~4.
Then, we see that they satisfy the relation (\ref{rel-smo-sepconst}).

Given commuting symmetry operators ${\cal S}^{(i)}$
for Eq.\ (\ref{originalSchEq}),
we perform the transformation
\be
 {\cal S}'^{(i)} = e^\Lambda {\cal S}^{(i)} e^{-\Lambda}
\ee
with an arbitrary function $\Lambda$.
Particularly, ${\cal S}'^{(0)}={\cal H}'$ [cf.\ (\ref{def_Hprime})].
Then we obtain
\be
 \left[{\cal S}'^{(i)},{\cal S}'^{(j)}\right]
 = e^\Lambda \left[{\cal S}^{(i)},{\cal S}^{(j)}\right] e^{-\Lambda} = 0 \,,
 \qquad i,j=0,1,2,\dots
\ee
and hence ${\cal S}'^{(i)}$ are commuting symmetry operators
for Eq.\ (\ref{Hprimeeq}). 
The existence of commuting symmetry operators is preserved
under the gauge transformation (\ref{gauge_transf_SchEq}).

\subsection{Construction of first- and second-order symmetry operators}
\label{subsec:construct-sop}

In the Eisenhart-Duval lift, the symmetry operators 
for the equation of motion (\ref{originalSchEq}) 
can be constructed from the symmetry operators for the massless Klein-Gordon equation (\ref{liftedLaplaceEq}).
To show it, we consider first- and second-order differential operators on $\tilde{M}$ in the covariant forms,\footnote{
We abbreviate ``$(i)$'' indexing Killing tensors $K^{(i)}$ in this subsection.}
\begin{equation}
 \tilde{{\cal L}} = \tilde{L}^A\tilde{\nabla}_A \,, \qquad
 \tilde{{\cal K}} = \tilde{\nabla}_A \tilde{K}^{AB} \tilde{\nabla}_B \,,
 \label{symoponMtilde}
\end{equation}
where $\tilde{\bm{L}}=\tilde{L}^A\partial_A$ is a vector
and $\tilde{\bm{K}}=\tilde{K}_{AB}d\tilde{x}^Ad\tilde{x}^B$ is a rank-2 symmetric tensor on $\tilde{M}$.
The conditions that $\tilde{{\cal L}}$ and $\tilde{{\cal K}}$ become
the symmetry operators for Eq.\ (\ref{liftedLaplaceEq}) are
\be
 \left[\tilde{\Box},\tilde{{\cal L}}\right] = 0 \,, \qquad
 \left[\tilde{\Box},\tilde{{\cal K}}\right] = 0 \,,
 \label{commutator_boxK}
\ee
which give rise to the conditions
\be
 \tilde{\nabla}_{(A}\tilde{L}_{B)} = 0 \,, \qquad
 \tilde{\nabla}_{(A}\tilde{K}_{BC)} = 0 \,.
 \label{Killingtensorcond}
\ee
These conditions mean that $\tilde{\bm{L}}$ and $\tilde{\bm{K}}$ are 
a Killing vector and a Killing tensor on $(\tilde{M},\tilde{\bm{g}})$, respectively.
Moreover, we obtain the anomaly-free condition \cite{Carter:1977pq},
\be
 \tilde{\nabla}^A \left(\tilde{K}_{[A}{}^C\tilde{R}_{B]C}\right) = 0 \,,
 \label{anomalyfreecond}
\ee
where $\tilde{R}_{AB}$ is the components of the Ricci tensor 
$\tilde{\bm{Ric}}=\tilde{R}_{AB}d\tilde{x}^Ad\tilde{x}^B$
on $(\tilde{M},\tilde{\bm{g}})$.
Thus we find that, if a Killing vector $\tilde{\bm{L}}$ is given, 
we can obtain the symmetry operator $\tilde{{\cal L}}$ by (\ref{symoponMtilde});
on the other hand, even if a Killing tensor $\tilde{\bm{K}}$ is given,
we cannot obtain the symmetry operator $\tilde{{\cal K}}$ by (\ref{symoponMtilde}) necessarily.
To construct the symmetry operator $\tilde{{\cal K}}$,
the anomaly-free condition (\ref{anomalyfreecond}) should be satisfied
for a given Killing tensor $\tilde{\bm{K}}$.

Some Killing tensors $\tilde{\bm{K}}$ are given as the square of 
a rank-$p$ Killing-Yano tensor $\tilde{\bm{f}} = (1/p!) \tilde f_{A_1A_2\dots A_p}
d\tilde{x}^{A_1}\wedge d\tilde{x}^{A_2}\wedge \dots \wedge d\tilde{x}^{A_p}$ by 
$\tilde K_{AB}= \tilde f_{AC_1\dots C_{p-1}}\tilde f_B{}^{C_1\dots C_{p-1}}$. 
For such Killing tensors the anomaly-free condition 
(\ref{anomalyfreecond}) is automatically satisfied 
since $\tilde K_{[A}{}^C \tilde R_{B]C}$ identically vanishes.
Contrarily, for Killing tensors that cannot be expressed 
in terms of a Killing-Yano tensor, $\tilde K_{[A}{}^C \tilde R_{B]C}$
 does not vanish in general, and in such a case the anomaly-free 
 condition (\ref{anomalyfreecond}) 
 must be satisfied only after taking the divergence $\tilde \nabla^A$.
Later we see that the Killing tensors 
associated with the symmetry operators for the Teukolsky equation and 
the LFKK equation fall into the latter case.

There are various ways to construct Killing vectors and tensors, and
one obvious way is to solve the Killing vector and tensor equations
 (\ref{Killingtensorcond}) directly.
A more sophisticated method is,
for example, to utilize the integrability conditions~\cite{Houri:2017tlk}. 
In this paper, we use the method proposed by Benenti \cite{MR1156532} that provided the canonical form of metrics admitting the separability of the geodesic equations and Killing tensors associated to them.

Now, we suppose that
the components of the Killing vector $\tilde{\bm{L}}$
and the Killing tensor $\tilde{\bm{K}}$ on $\tilde{M}$
are independent of the coordinates $u$ and $v$,
\be
 \partial_u \tilde{L}^A=0 \,, \qquad
 \partial_v \tilde{L}^A=0 \,, \qquad
 \partial_u \tilde{K}^{AB}=0 \,, \qquad
 \partial_v \tilde{K}^{AB}=0 \,.
 \label{compindepuv}
\ee
Applying the symmetry operators $\tilde{{\cal L}}$ and
$\tilde{{\cal K}}$ on $\tilde{M}$, defined by by (\ref{symoponMtilde}),
to the wave function $\tilde \Phi$ of the form (\ref{restrictedformofsolutions})
together with $E=0$,
we obtain the first and second-order differential operators ${\cal L}$ and ${\cal K}$ on $M$ by
\begin{equation}
 \tilde{{\cal L}}\tilde \Phi = e^{iv} {\cal L}\Phi \,, \qquad
 \tilde{{\cal K}}\tilde \Phi = e^{iv} {\cal K}\Phi \,.
    \label{symoponM}
\end{equation}
Since a simple calculation gives
\be
 [\tilde{\Box}, \tilde{{\cal L}}]\tilde \Phi = e^{iv} [{\cal H},{\cal L}]\Phi \,, 
 \qquad
 [\tilde{\Box}, \tilde{{\cal K}}]\tilde \Phi = e^{iv} [{\cal H},{\cal K}]\Phi \,,
\ee
the conditions (\ref{commutator_boxK}) are satisfied if and only if
\begin{equation}
 [{\cal H},{\cal L}]=0 \,, \qquad
 [{\cal H},{\cal K}]=0 \,.
\end{equation}
Thus we have found that, starting from a Killing vector $\tilde{\bm{L}}$
and a Killing tensor $\tilde{\bm{K}}$ on $\tilde{M}$, 
we can construct the first- and second-order symmetry operators 
${\cal L}$ and ${\cal K}$ for the equation of motion (\ref{originalSchEq}).

To express the forms of ${\cal L}$ and ${\cal K}$ explicitly,
we recast the components of the Killing vector $\tilde{\bm{L}}$ 
and Killing tensor $\tilde{\bm{K}}$ on $\tilde{M}$ as
\begin{align}\label{killing1_mod}
 \left(\tilde{L}^A\right)
 &= \left(\tilde{L}^a,\tilde{L}^u,\tilde{L}^v\right)
 = \Big(L^a,I,J-qL^aA_a\Big) \,, \\
  \left(\tilde{K}^{AB}\right)&=
\left(\begin{array}{ccc}
\tilde{K}^{ab} & \tilde{K}^{a u} & \tilde{K}^{a v}\\
\tilde{K}^{u b} & \tilde{K}^{uu} & \tilde{K}^{uv}\\
\tilde{K}^{vb} & \tilde{K}^{vu} & \tilde{K}^{vv}
\end{array}\right) \nonumber\\
&=
\left(\begin{array}{ccc}
K^{ab} & U^a &
 N^a-q K^{ac} A_c
 \\
U^b & C & T + 2 CV-q U^c A_c
\\
N^b-q K^{bc} A_c 
&
T + 2CV-q U^c A_c
&
W - 2 q N^c A_c + q^2 K^{cd}A_c A_d
\end{array}\right)\,,
\label{tildeK_components}
\end{align}
where $\tilde{L}^A$ and $\tilde{K}^{AB}$ have been recast 
into $(L^a, I, J)$ and $(K^{ab}, U^a, N^a, C, T, W)$ without loss of generality.
Here, using the fact that 
$\tilde L^{A}$ is a Killing vector and 
$\tilde K^{AB}$ is a Killing tensor on $\tilde M$,
we can show that
$L^a$ and $U^a$ are Killing vectors and
$K^{ab}$ is a Killing tensor on $M$, 
$I$ and $C$ are constants,
and also
$\nabla_a N^a = 0$ (see appendix \ref{app:Killingeq}).
Using the new variables,
the symmetry operators ${\cal L}$ and ${\cal K}$ are expressed as
\begin{align}
{\cal L}
&= L^a (\nabla_a-iqA_a) + iJ\,, \\
{\cal K}
&= \left(
\nabla_a - iqA_a
\right) K^{ab}
\left(
\nabla_b - iq A_b
\right)
+ 2i N^a (\nabla_a - iq A_a) - W
\, .
\label{symop_explicit}
\end{align}

\subsection{Commutativity conditions of the symmetry operators}
\label{sec:com-smo}

To apply the symmetry operators to the separation of variables, 
it is important that they commute with each other 
so that they generate independent separation constants.
In this subsection we examine the conditions 
necessary for such commutativity of the symmetry operators.

Below we focus on the symmetry operators given by (\ref{symoponMtilde}) 
and (\ref{commutator_boxK}).
The conditions for two symmetry operators to commute are
\begin{equation}
\left[
\tilde{\cal L}^{(i)},
\tilde{\cal L}^{(j)}
\right]  =
0 \,, \qquad
\left[
\tilde{\cal L}^{(i)},
\tilde{\cal K}^{(j)}
\right]  =
0 \,, \qquad
\left[
\tilde{\cal K}^{(i)},
\tilde{\cal K}^{(j)}
\right]  =
0
\,.
\label{KKcond_upstairs}
\end{equation}
These conditions were studied in detail by Kolar and Krtous 
\cite{Kolar:2015cha}, according to which the conditions are reduced to
the commutation relations of the Killing vector $\tilde{\bm{L}}$ 
and the Killing tensor $\tilde{\bm{K}}$
with respect to the Schouten-Nijenhuis bracket,
\begin{equation}
  \left[
  \tilde{\bm{L}}^{(i)},
  \tilde{\bm{L}}^{(j)}
  \right]_\text{SN}
  =0 \,, \qquad
    \left[
  \tilde{\bm{L}}^{(i)},
  \tilde{\bm{K}}^{(j)}
  \right]_\text{SN}
  =0 \,, \qquad
  \left[
  \tilde{\bm{K}}^{(i)},
  \tilde{\bm{K}}^{(j)}
  \right]_\text{SN}
  =0
  \, ,
  \label{KKcond1}
\end{equation}
where the Schouten-Nijenhuis brackets $[~,~]_\text{SN}$
are defined by
\begin{align}
&\left(\left[
  \tilde{\bm{L}}^{(i)},
  \tilde{\bm{L}}^{(j)}
  \right]_\text{SN}\right)_A
  \equiv
  \tilde L^{(i)B} \tilde \nabla_B \tilde L^{(j)}_A
  -
  \tilde L^{(j)B} \tilde \nabla_B \tilde L^{(i)}_A \, , \\
& \left(\left[
  \tilde{\bm{L}}^{(i)},
  \tilde{\bm{K}}^{(j)}
  \right]_\text{SN}\right)_{AB}
  \equiv
  \tilde L^{(i)C}\tilde \nabla_C \tilde K^{(j)}_{AB}
  -
  \tilde K^{(j)}_{D(A}\tilde \nabla^D \tilde L^{(i)}_{B)}\, , \\
& \left(\left[
  \tilde{\bm{K}}^{(i)},
  \tilde{\bm{K}}^{(j)}
  \right]_\text{SN}\right)_{ABC}
  \equiv
  \tilde K^{(i)}_{D(A}\tilde \nabla^D \tilde K^{(j)}_{BC)}
  -
  \tilde K^{(j)}_{D(A}\tilde \nabla^D \tilde K^{(i)}_{BC)} \,,
\end{align}
and the condition, 
which comes from the commutation relations of two Killing tensors,
\begin{equation}
\nabla^A \tilde m_{AB}^{(i,j)} = 0~,
\label{commutativity}
\end{equation}
where
\begin{equation}
\tilde m_{AB}^{(i,j)}
\equiv
\left(
\tilde K^{(i)}_{C[A}\tilde \nabla^D \tilde \nabla^C \tilde K^{(j)}_{B]D}
-
\tilde K^{(j)}_{C[A}\tilde \nabla^D \tilde \nabla^C \tilde K^{(i)}_{B]D}
\right)
-
\left(
\tilde \nabla^D \tilde K^{(i)}_{C[A}
\right)
\left(
\tilde \nabla^C \tilde K^{(j)}_{B]D}
\right)
- 3
\tilde K^{(i)}_{C[A} \tilde K^{(j)}_{B]D}
\tilde R^{CD}
\, .
\label{KKcond2}
\end{equation}
It is worth noting that the metric itself is a Killing tensor. 
If we set $\tilde{\bm{K}}=\tilde{\bm{g}}$ in (\ref{KKcond1}) 
and (\ref{commutativity}), 
we obtain the conditions (\ref{Killingtensorcond}) 
and (\ref{anomalyfreecond}) as a result.

Finally, we demonstrate the commutativity of the operators 
${\cal S}^{(i)}$ on $(M, \bm{g})$ constructed from the commuting symmetry 
operators $\tilde{\cal S}^{(i)}$ on $(\tilde M,\tilde{\bm{g}})$ by
\be
 \tilde{\cal S}^{(i)}\tilde \Phi = e^{iv}{\cal S}^{(i)}\Phi \,.
\ee
By a simple calculation, we obtain
\begin{equation}
  \left[
  \tilde{\cal S}^{(i)},
  \tilde{\cal S}^{(j)}
  \right]\tilde \Phi=
e^{iv}
\left[
{\cal S}^{(i)},
{\cal S}^{(j)}
\right]\Phi
\,.
\end{equation}
It follows consequently that $\tilde{\cal L}^{(i)}$ and $\tilde{\cal K}^{(j)}$
on the uplifted spacetime 
$(\tilde M,\tilde g_{AB})$
commute with each other even on the original spacetime $(M, g_{ab})$ 
once the conditions (\ref{KKcond1}) and (\ref{commutativity}) are satisfied,
and we obtain 
\begin{equation}
\left[
{\cal L}^{(i)},
{\cal L}^{(j)}
\right]  =
0 \,, \qquad
\left[
{\cal L}^{(i)},
{\cal K}^{(j)}
\right]  =
0 \,, \qquad
\left[
{\cal K}^{(i)},
{\cal K}^{(j)}
\right]  =
0
\,,
\label{KKcond_downstairs}
\end{equation}
where ${\cal L}^{(i)}$ and ${\cal K}^{(i)}$ are the ones obtained by (\ref{symoponM}).
As explained at Eq.~(\ref{rel-smo-sepconst}), the eigenvalues of these commuting symmetry operators,
\begin{equation}
{\cal K}^{(i)} \Phi = \kappa_i \Phi~,
\qquad
{\cal L}^{(i)} \Phi = i \omega_k \Phi~,
\label{separationconstants}
\end{equation}
become the separation constants for the equation of motion (\ref{originalSchEq}), which is expressed as ${\cal K}^{(0)} \Phi = E\, \Phi$.

\vspace{0.5cm}
\begin{figure}[t]
\begin{center}
\includegraphics[scale=0.5]{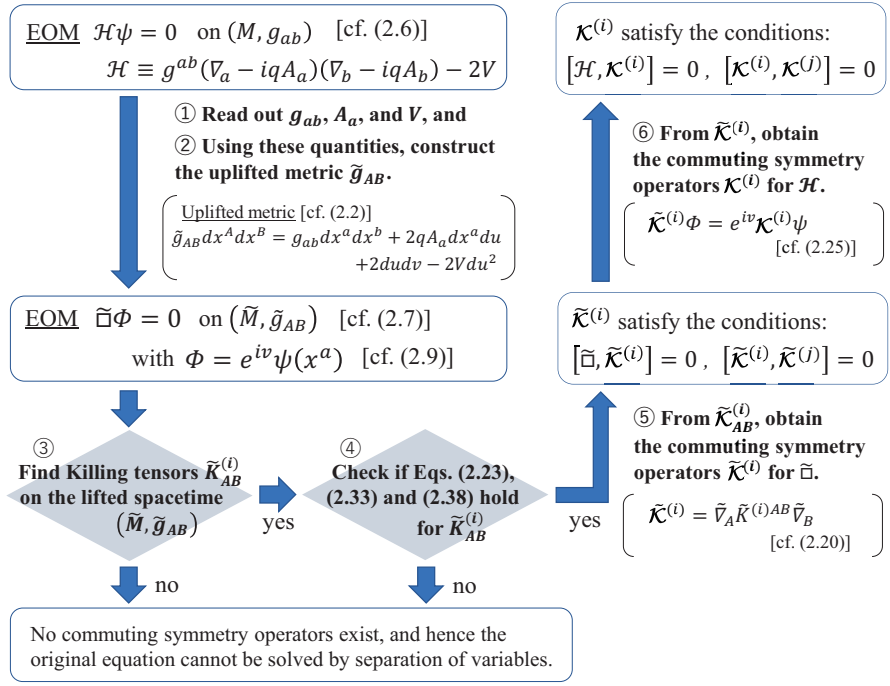}
\caption{The procedure for obtaining the commuting symmetry operators ${\cal K}^{(i)}$ 
for the equation of motion (\ref{originalSchEq}).
In this procedure, we must check if there exist Killing tensors $\tilde{\bm{K}}^{(i)}$ 
on the uplifted spacetime $(\tilde{M},\tilde{\bm{g}})$ 
that satisfy the conditions (\ref{anomalyfreecond}),
(\ref{KKcond1}) and (\ref{KKcond2})
at the steps 3 and 4.
First-order symmetry operator ${\cal L}^{(i)}$ can be constructed by a similar procedure.
}
\end{center}
\end{figure}

To summarize, the procedure for obtaining the commuting symmetry operators ${\cal K}^{(i)}$ 
for the equation of motion (\ref{originalSchEq}) is given as follows.
First-order symmetry operator ${\cal L}^{(i)}$ can be constructed in a similar manner.
\begin{enumerate}
 \item Given the equation of motion in the form (\ref{originalSchEq}),
 read out $\bm{g}$, $\bm{A}$ and $V$ from its coefficients.
 \item Using these quantities,
  construct the uplifted metric $\tilde{\bm{g}}$ as (\ref{liftedmetric}).
 \item Find Killing tensors $\tilde{\bm{K}}^{(i)}$ on ($\tilde{M},\tilde{\bm{g}}$)
 such that the components are independent of $u$ and $v$.
 \item For the Killing tensors found in the step 3,
 check if the anomaly-free condition (\ref{anomalyfreecond}),
and also the commutativity conditions (\ref{KKcond1}) and (\ref{KKcond2})
  hold.
 \item If they hold, construct the commuting symmetry operators $\tilde{{\cal K}}^{(i)}$
 by (\ref{symoponMtilde}).
 Then it follows that $[\tilde{\Box},\tilde{{\cal K}}^{(i)}]=0 = [\tilde{{\cal K}}^{(i)},\tilde{{\cal K}}^{(j)}]$.
 \item Finally, obtain commuting symmetry operators ${\cal K}^{(i)}$ from $\tilde{{\cal K}}^{(i)}$
 via (\ref{symoponM}) and (\ref{symop_explicit}). Then $[{\cal H},{\cal K}^{(i)}]=0=[{\cal K}^{(i)},{\cal K}^{(j)}]$.
\end{enumerate}

\section{Symmetry operators for the Teukolsky equation}
\label{sec:sec3}

In the previous section, we summarized the procedure 
to construct commuting symmetry operators 
for the equation of motion (\ref{originalSchEq}).
The aim of this section is to apply this procedure to the Teukolsky equation
on the four-dimensional Kerr-NUT-(A)dS spacetime
to construct its symmetry operator.
To this end, we first recall the Carter form \cite{Carter:1968ks} 
of the four-dimensional Kerr-NUT-(A)dS metric
in Sec.~\ref{sec:Carter4D} and explain Benenti's method 
for constructing Killing tensors~\cite{Benenti1979}
by explicitly constructing the Killing tensor 
on the Kerr-NUT-(A)dS spacetime in Sec.~\ref{sec:KTonKNAdS}.
Benenti's method is used repeatedly in subsequent sections.
After that, we review the separation of variables 
in the Teukolsky equation \cite{PhysRevLett.29.1114,Teukolsky:1973ha}
briefly in Sec.~\ref{sec:sop-Teukolsky} (see appendix B for details)
and then construct the symmetry operator for the Teukolsky equation
in Sec.~\ref{sec:Benn+}.

\subsection{Carter form of the Kerr-NUT-(A)dS metric}
\label{sec:Carter4D}

The Kerr metric in the Boyer-Lindquist coordinates $(t,r,\theta,\phi)$ is given by
\be
 ds^2 = -\frac{\Delta}{\Sigma}\left(dt - a\sin^2\theta d\phi\right)^2
 + \frac{\Sigma}{\Delta}dr^2 + \Sigma d\theta^2
  + \frac{\sin^2\theta}{\Sigma}\left(adt - (r^2+a^2)d\phi\right)^2 \,,
\ee
where
\be
 \Delta = r^2 -2mr+a^2 \,, \qquad
 \Sigma = r^2 + a^2\cos^2\theta \,.
\ee
This metric describes a rotating black hole in vacuum with mass $m$
and angular momenta $J=ma$.
Performing the coordinate transformation
\be
 p = -a \cos\theta \,, \quad
 \tau = t - a\phi
 \,, \quad
 \sigma = -\frac{\phi}{a}
 \,,
\ee
the metric is written in the Carter form \cite{Carter:1968ks}
\begin{eqnarray}
ds^2=-\frac{{\cal Q}(r)}{r^2+p^2}(d\tau-p^2 d\sigma)^2
+\frac{r^2+p^2}{{\cal Q}(r)}dr^2
 +\frac{r^2+p^2}{{\cal P}(p)}dp^2
+\frac{{\cal P}(p)}{r^2+p^2}(d\tau+r^2 d\sigma)^2 \,,
\label{Cartermetric}
\end{eqnarray}
where
\be
 {\cal Q}(r) = r^2 -2 mr + a^2 \,, \quad
 {\cal P}(p) = a^2 - p^2 \,.
 \label{funcQandP}
\ee
The metric (\ref{Cartermetric}) satisfies the vacuum Einstein equation
with cosmological constant $\lambda$,
\be
 R_{ab}=\lambda g_{ab} \,,
 \label{Eeq}
\ee
if and only if ${\cal Q}(r)$ and ${\cal P}(p)$ are given by
\be
 {\cal Q}(r) = -\frac{\lambda}{3}r^4 + \epsilon r^2
 -2 mr + a^2 \,, \quad
 {\cal P}(p) = -\frac{\lambda}{3}p^4 - \epsilon p^2
 +2 lp + a^2 \,,
 \label{funcQandP_Einstein}
\ee
with constants $\epsilon$, $a$, $m$ and $l$.
This solution is called the Kerr-NUT-(A)dS metric.
Once ${\cal Q}(r)$ and ${\cal P}(p)$ are replaced with arbitrary functions 
depending only on $r$ and $p$ respectively, the metric (\ref{Cartermetric}) 
does not satisfy the Einstein equation~(\ref{Eeq}) anymore, 
but it still admits the Killing tensor as explained in the next section. 
In this paper, we work on the metric (\ref{Cartermetric}) 
with arbitrary ${\cal Q}(r)$ and ${\cal P}(p)$, 
which is called the off-shell Kerr-NUT-(A)dS metric
or simply the Carter metric.
The spacetime described by the (off-shell) Kerr-NUT-(A)dS metric
is called the (off-shell) Kerr-NUT-(A)dS spacetime.

\subsection{Killing tensor on the Kerr-NUT-(A)dS spacetime}
\label{sec:KTonKNAdS}

The (off-shell) Kerr-NUT-(A)dS spacetime admits a nontrivial Killing tensor, 
which guarantees the complete integrability and separability 
of the geodesic equation \cite{MR1156532}.
To provide the Killing tensor explicitly, Benenti's method is useful.

Benenti's method is the following:
We first denote local coordinates $(x^a)$ by $(x_\mu,\psi_k)$,
where $\psi_k$ are the Killing coordinates of a $D$-dimensional metric $\bm{g}$, i.e.
the metric components $g^{ab}$ do not depend on these coordinates.
When the number of the Killing coordinates $\psi_k$ is $D-n$,
the Greek indices $\mu,\nu,\dots$ take $0,1,\dots,n-1$
and the Latin indices $k,\ell,\dots$ take $n,n+1,\dots,D-1$.
If the components of the inverse metric $g^{ab}$
are written in Benenti's canonical form
\begin{equation}
 g^{\mu\mu} = \bar{\phi}^{\mu+1}{}_{(1)} \,,\qquad
 g^{\mu a} = 0 \quad (a\neq \mu) \,, \qquad
 g^{k\ell} = \sum_{\mu=0}^{n-1}  \zeta_{(\mu)}^{k\ell}\bar{\phi}^{\mu+1}{}_{(1)} \,,
 \label{BenentiCanonicalFormMetric}
\end{equation}
where $\bar{\phi}^\mu{}_{(\nu)}$ is the element
of the inverse St\"ackel matrix $\bar{\phi}$
and $\zeta^{k\ell}_{(\mu)}$ is the element of the $\mu$th $\zeta$-matrix $\zeta_{(\mu)}$,
one can construct $n$ Killing tensors $\bm{K}_{(\nu)}$
whose contravariant components $K^{ab}_{(\nu)}$ are given by
\begin{equation}
 K^{\mu\mu}_{(\nu)} = \bar{\phi}^{\mu+1}{}_{(\nu+1)} \,,\qquad
 K^{\mu a}_{(\nu)} = 0 \quad (a\neq \mu) \,, \qquad
 K^{k\ell}_{(\nu)} = \sum_{\mu=0}^{n-1}  \zeta_{(\mu)}^{k\ell}\bar{\phi}^{\mu+1}{}_{(\nu+1)} \,.
 \label{BenentiCanonicalFormKT}
\end{equation}
Since we have $\bm{K}_{(0)}=\bm{g}$,
we obtain $n-1$ nontrivial Killing tensors.
Here, the inverse St\"ackel matrix is the inverse of the St\"ackel matrix, i.e., $\bar{\phi}\phi=1$,
and the element $\phi^{(\mu)}{}_\nu$ of the St\"ackel matrix $\phi$
must depend only on the coordinate $x^\nu$,
and any element $\zeta^{k\ell}_{(\mu)}$ of the $\mu$th $\zeta$-matrix $\zeta_{(\mu)}$
must depend only on the coordinate $x_\mu$.

To illustrate it, 
we shall construct the Killing tensor on the off-shell Kerr-NUT-(A)dS spacetime.
For the off-shell Kerr-NUT-(A)dS metric (\ref{Cartermetric}), 
the components of the inverse metric are given by
\begin{equation}
\begin{aligned}
& g^{rr}= \frac{{\cal Q}}{r^2+p^2} \,, \qquad
  g^{pp}= \frac{{\cal P}}{r^2+p^2} \,, \\
& g^{\tau \tau} = -\frac{r^4}{{\cal Q}^2}g^{rr}
  +\frac{p^4}{{\cal P}^2}g^{pp} \,, \qquad
  g^{\tau \sigma} = \frac{r^2}{{\cal Q}^2}g^{rr}
  +\frac{p^2}{{\cal P}^2}g^{pp} \,, \qquad
  g^{\sigma \sigma} = - \frac{1}{{\cal Q}^2}g^{rr}
  +\frac{1}{{\cal P}^2}g^{pp} \,.
\end{aligned}
\label{g-Carter4D}
\end{equation}
They fit into Benenti's canonical form (\ref{BenentiCanonicalFormMetric}) when 
the St\"{a}ckel matrix $\phi$ and its inverse $\bar{\phi}$ are given by
{\renewcommand{\arraystretch}{2}
\begin{equation}
\phi = \left(
\begin{array}{cc}
\displaystyle{\frac{r^2}{{\cal Q}} }&\displaystyle{\frac{p^2}{{\cal P}} }
\\
\displaystyle{\frac{1}{{\cal Q}}} &-\displaystyle{\frac{1}{{\cal P}}}  
\end{array}
\right)
\,, \quad
 \bar{\phi} = \left(
\begin{array}{cc}
\displaystyle{\frac{{\cal Q}}{r^2+p^2}} & \displaystyle{\frac{p^2{\cal Q}}{r^2+p^2}}  \\
\displaystyle{\frac{{\cal P}}{r^2+p^2}} & \displaystyle{-\frac{r^2{\cal P}}{r^2+p^2}} 
\end{array}
\right)\,, 
\end{equation}}%
and $\zeta$-matrices $\zeta_{(\mu)}$ are given by
\begin{equation}
\zeta_{(r)} = \frac{1}{{\cal Q}^2}
 \left(
\begin{array}{cc}
-r^4 &r^2  \\
r^2 &-1 
\end{array}
\right)\,, \quad
 \zeta_{(p)} = \frac{1}{{\cal P}^2}
 \left(
\begin{array}{cc}
p^4 &p^2  \\
p^2 &1 
\end{array}
\right)\,.
\end{equation}
Here, we have set $n=2$. Namely,
the Greek indices $\mu,\nu,\dots$ take $0,1$ (or $r,p$)
and the Latin indices $k,\ell,\dots$ take $2,3$ (or $\tau,\sigma$),
and hence we have  $(x^a)=(x_0,x_1,\psi_2,\psi_3)=(r,p,\tau,\sigma)$.
From (\ref{BenentiCanonicalFormKT}), 
the contravariant components of the nontrivial Killing tensor
are obtained by replacing $\bar{\phi}^\mu{}_{(1)}$ 
with $\bar{\phi}^\mu{}_{(2)}$ as
\begin{equation}
\begin{aligned}
& K^{rr}= \frac{p^2{\cal Q}}{r^2+p^2} \,, \qquad
  K^{pp}= -\frac{r^2{\cal P}}{r^2+p^2} \,, \\
& K^{\tau \tau} = -\frac{r^4}{{\cal Q}^2}K^{rr}
  +\frac{p^4}{{\cal P}^2}K^{pp} \,, \qquad
  K^{\tau \sigma} = \frac{r^2}{{\cal Q}^2}K^{rr}
  +\frac{p^2}{{\cal P}^2}K^{pp} \,, \qquad
  K^{\sigma \sigma} = - \frac{1}{{\cal Q}^2}K^{rr}
  +\frac{1}{{\cal P}^2}K^{pp} \,.
\end{aligned}
\label{K-Carter4D}
\end{equation}
Thus we obtain the Killing tensor
\begin{equation}
 \bm{K}
 =p^2\left(-\frac{{\cal Q}(r)}{r^2+p^2}(d\tau-p^2 d\sigma)^2
+\frac{r^2+p^2}{{\cal Q}(r)}dr^2\right)
 - r^2\left(\frac{r^2+p^2}{{\cal P}(p)}dp^2
+\frac{{\cal P}(p)}{r^2+p^2}(d\tau+r^2 d\sigma)^2\right) \,.
     \label{KillingtensoronKerr}
\end{equation}

To express the Killing tensor in a simple form, 
it is convenient to introduce the orthonormal basis of 1-forms
\begin{equation}
\bm{e}^{\underline 0} = \sqrt\frac{{\cal Q}(r)}{r^2+p^2} \left(d\tau - p^2 d\sigma \right)~,
~~
\bm{e}^{\underline 1} = \sqrt\frac{r^2+p^2}{{\cal Q}(r)} dr~,
~~
\bm{e}^{\underline 2} = \sqrt\frac{r^2+p^2}{{\cal P}(p)} dp~,
~~
\bm{e}^{\underline 3} = \sqrt\frac{{\cal P}(p)}{r^2+p^2} \left(d\tau + r^2 d\sigma \right)~,
\label{orthbasis1-form}
\end{equation}
with which the metric (\ref{Cartermetric}) is expressed as
\begin{equation}
\bm{g} = 
-\bm{e}^{\underline 0} \bm{e}^{\underline 0} 
+ \bm{e}^{\underline 1} \bm{e}^{\underline 1}
+ \bm{e}^{\underline 2} \bm{e}^{\underline 2}
+ \bm{e}^{\underline 3} \bm{e}^{\underline 3} \,.
\end{equation}
We also introduce the dual vector basis by $\bm{e}^{\underline a}(\bm{e}_{\underline b})=\delta^{\underline a}{}_{\underline b}$.
The vector basis dual to (\ref{orthbasis1-form}) are given by
\begin{equation}
\bm{e}_{\underline 0} = \frac{1}{\sqrt{{\cal Q}(r)\left(r^2+p^2\right)}}
\left(
r^2 \partial_\tau - \partial_\sigma
\right)
~,
~~
\bm{e}_{\underline 1} = \sqrt\frac{{\cal Q}(r)}{r^2+p^2} \partial_r ~,
~~
\bm{e}_{\underline 2} = \sqrt\frac{{\cal P}(p)}{r^2+p^2} \partial_p ~,
~~
\bm{e}_{\underline 3} = 
\frac1{
\sqrt{{\cal P}(p)\left(r^2+p^2\right)}
} 
\left(
p^2 \partial_\tau + \partial_\sigma
\right)~.
\end{equation}
The Killing tensor is written as
\begin{equation}
\bm{K} = 
p^2\Bigl(-\bm{e}^{\underline 0} \bm{e}^{\underline 0} 
+ \bm{e}^{\underline 1} \bm{e}^{\underline 1}\Bigr)
- r^2\Bigl(\bm{e}^{\underline 2} \bm{e}^{\underline 2}
+ \bm{e}^{\underline 3} \bm{e}^{\underline 3}\Bigr) \,.
\end{equation}

\subsection{Separation of variables in the Teukolsky equation}
\label{sec:sop-Teukolsky}

When a spacetime admits a gauged conformal Killing-Yano tensor (GCKY),
the Maxwell equation (\ref{Maxwellequation}) on such a spacetime can reduce
to a scalar-type equation by the method of the Hertz potential
(see appendix~\ref{App:Teukolsky} for details).
This scalar-type equation is not always solvable by separation of variables,
but the Kerr spacetime is known to admit three GCKYs and hence
the Maxwell equation reduces to three scalar-type equations.
An interesting thing is that two of them coincide with the Teukolsky equation for $s=\pm 1$
\cite{PhysRevLett.29.1114,Teukolsky:1973ha},
which was proposed as the master equation for perturbations of Maxwell fields
 on the Kerr spacetime, and can be solved by separation of variables.

Since even the off-shell Kerr-NUT-(A)dS metric (\ref{Cartermetric}) admits three GCKYs, 
the Maxwell equation reduces to three scalar-type equations.
Two of them can be written in the form
\begin{align}\label{Teq_0}
 {\cal H}\Phi \equiv \Big[g^{ab}(\nabla_a + s A_a)(\nabla_b+sA_b) 
 - \frac{R}{6}- 4 s^2 \Psi_2\Big]\Phi = 0 \,,\qquad
 s=\pm 1 \,,
\end{align}
where $R$ and $\Psi_2$ are the scalar curvature and the Weyl scalar, given by
\begin{align}
 R = -\frac{{\cal Q}''+{\cal P}''}{r^2+p^2} \,, \qquad
\Psi_2 = \frac{r^2+p^2}{12}\Big[\partial_r\Big(\frac{\partial_r Q_T}{r^2+p^2}\Big)
-\partial_p\Big(\frac{\partial_p Q_T}{r^2+p^2}\Big)\Big]
 -\frac{i}{4}\partial_r \partial_p Q_T
\end{align}
with $Q_T=({\cal Q}(r)-{\cal P}(p))/(r^2+p^2)$,
and the gauge potential ${\bm A}$ is given by
\begin{align}
&\bm{A}
 = \frac{{\cal Q}'}{2{\cal Q}}dr
 - \frac{ \chi {\cal Q}'-4{\cal Q} }{ 2\chi^2\bar{\chi} }(d\tau-p^2d\sigma)
 - \frac{ i\chi {\cal P}'+4{\cal P} }{ 2\chi^2\bar{\chi} }(d\tau+r^2d\sigma)
 \label{ATVT}
\end{align}
with $\chi=r+ip$ and $\bar{\chi}=r-ip$.
This equation coincides with the Teukolsky equation with $s=\pm1$
(actually also with $s=0$$, \pm 1/2$ and $\pm 2$)
when the off-shell Kerr-NUT-(A)dS metric is restricted to the Kerr metric,
i.e., ${\cal Q}$ and ${\cal P}$ are given by (\ref{funcQandP}).
In what follows, we still call Eq.~(\ref{Teq_0}) with arbitrary $s$
the Teukolsky equation.
The operator ${\cal H}$ of Eq.~\eqref{Teq_0} is explicitly given by
\begin{align}\label{Teq}
 {\cal H}&=
 \frac{1}{r^2+p^2}\Bigg[{\cal Q}\partial_r^2 +(1+s){\cal Q}'\partial_r
 +{\cal P}\partial_p^2
 + {\cal P}'\partial_p
 - \frac{1}{{\cal Q}}(r^2\partial_\tau -\partial_\sigma)^2
 + \frac{1}{{\cal P}}(p^2\partial_\tau +\partial_\sigma)^2 \nonumber\\
 & + s \Bigg(\frac{{\cal Q}'}{{\cal Q}}\left(r^2\partial_\tau - \partial_\sigma\right)
 - \frac{i{\cal P}'}{{\cal P}}\left(p^2\partial_\tau + \partial_\sigma\right)
  - 4 (r-ip)\partial_\tau\Bigg)
  + \frac{(s+1)(2s+1)}{6}{\cal Q}''
   +\frac{(2s^2+1)}{6}{\cal P}''
   - s^2\frac{{\cal P}'^2}{4{\cal P}}\Bigg] \,.
\end{align}

We immediately find that this equation can be solved by separation of variables
by setting
\be
 \Phi = e^{i\omega \tau}e^{im\sigma} R(r)\Theta(p) \,,
 \label{sepAnsatz}
\ee
where $\omega$ and $m$ are constants.
Thus it results in the separated equations in terms of $r$ and $p$ given by
\begin{align}
& \frac{1}{{\cal Q}^s}\frac{d}{dr}\left({\cal Q}^{s+1}\frac{dR}{dr}\right)
 + \left[\frac{i s (\omega r^2-m){\cal Q}'+(\omega r^2-m)^2}{{\cal Q}}
 + \frac{(s+1)(2s+1)}{6}{\cal Q}'' - 4is\omega r - \kappa \right]R = 0 \,,
 \label{sepEquation_Kerr1}\\
& \frac{d}{dp}\left({\cal P}\frac{d\Theta}{dp}\right)
 + \left[ - \frac{(s{\cal P}'-2\omega p^2-2m)^2}{4{\cal P}}
 + \frac{(2s^2+1)}{6} {\cal P}''
 +4s\omega p 
+ \kappa \right]\Theta = 0 \,,
 \label{sepEquation_Kerr2}
\end{align}
where $\kappa$ is a separation constant.

\subsection{Symmetry operators for the Teukolsky equation}
\label{sec:Benn+}

To explain the separability of the Teukolsky equation, 
we shall construct the symmetry operator for the Teukolsky equation.
To obtain the symmetry operator for Eq.~(\ref{Teq_0}),
we consider the uplifted metric,%
\footnote{The components of the uplifted metric $\tg_{AB}$ become complex,
since the vector potential used to lift the metric has complex components.
This is caused since
the gauge transformation of
the gauged conformal Killing-Yano tensor is $\mathbb{C}^*$, 
rather than $U(1)$ for the Maxwell field.
See appendix \ref{App:Teukolsky} for more details.}
which is constructed by Eq.~(\ref{liftedmetric}) 
with (\ref{Cartermetric}) and (\ref{ATVT}).
The components of the inverse uplifted metric $\tg^{AB}$ are given by
\begin{equation}
\begin{aligned}
 \tg^{ab} &= g^{ab} \,, \\
 \tg^{rv} &= -is \frac{{\cal Q}'}{2{\cal Q}} g^{rr} \,, \\
 \tg^{\tau v} &= -is \left( \frac{r^2{\cal Q}'}{2{\cal Q}^2}-\frac{2r}{{\cal Q}} \right)g^{rr}
 -s\left( \frac{p^2{\cal P}'}{2{\cal P}^2}-\frac{2p}{{\cal P}} \right)g^{pp} \,, \\
 \tg^{\sigma v} &= \frac{is{\cal Q}'}{2{\cal Q}^2}g^{rr}
 - \frac{s{\cal P}'}{2{\cal P}^2}g^{pp} \,, \\
 \tg^{uv} &=~ 1~=~ \frac{r^2}{{\cal Q}}g^{rr}+\frac{p^2}{{\cal P}}g^{pp} \,, \\
 \tg^{vv} &= - \frac{(2s^2+1) {\cal Q}''}{6{\cal Q}}g^{rr}
 -\left( \frac{(2s^2+1){\cal P}''}{6{\cal P}}-\frac{{\cal P}'^2}{4{\cal P}^2}\right)g^{pp} \,,
\end{aligned}
\label{lifted-g_Teukolsky}
\end{equation}
where $g^{ab}$ are given by (\ref{g-Carter4D}) 
and the components not listed here are vanishing.
It turns out that these components fit into Benenti's canonical form (\ref{BenentiCanonicalFormMetric}).%
\footnote{Precisely speaking, the $(r,v)$ component $g^{rv}$ is not fitting 
into Benenti's canonical form (\ref{BenentiCanonicalFormMetric}),
but generalizing the $\zeta$-matrices appropriately one can transform it 
so as to fit into Benenti's canonical form.
Then, one can construct a Killing tensor by (\ref{BenentiCanonicalFormKT})
and making the inverse transformation one obtains the $(r,v)$ component $K^{rv}$
of the Killing tensor. This result coincides with (\ref{lifted-K_Teukolsky}).}
We note that this property is a special feature of the Teukolsky equation
in the sense that the uplifted metric constructed
from a generic equation of motion is not guaranteed to fit into Benenti's canonical form.
From (\ref{BenentiCanonicalFormKT}),
the components of the Killing tensor are given by
\begin{equation}
\begin{aligned}
 \tilde{K}^{ab} &= K^{ab} \,, \\
 \tilde{K}^{rv} &= -is \frac{{\cal Q}'}{2{\cal Q}} K^{rr} \,, \\
 \tilde{K}^{\tau v} &= -is \left( \frac{r^2{\cal Q}'}{2{\cal Q}^2}-\frac{2r}{{\cal Q}} \right)K^{rr}
 -s\left( \frac{p^2{\cal P}'}{2{\cal P}^2}-\frac{2p}{{\cal P}} \right)K^{pp} \,, \\
 \tilde{K}^{\sigma v} &= \frac{is{\cal Q}'}{2{\cal Q}^2}K^{rr}
 - \frac{s{\cal P}'}{2{\cal P}^2}K^{pp} \,, \\
 \tilde{K}^{uv} &=~ 0~=~ \frac{r^2}{{\cal Q}}K^{rr}+\frac{p^2}{{\cal P}}K^{pp} \,, \\
 \tilde{K}^{vv} &= - \frac{(2s^2+1) {\cal Q}''}{6{\cal Q}}K^{rr}
 -\left( \frac{(2s^2+1){\cal P}''}{6{\cal P}}-\frac{{\cal P}'^2}{4{\cal P}^2}\right)K^{pp} \,,
\end{aligned}
\label{lifted-K_Teukolsky}
\end{equation}
and the other components are zero,
and $K^{ab}$ are given by (\ref{KillingtensoronKerr}).
Thus we obtain the symmetry operator of the Teukolsky equation,
\be
 {\cal K}
 = (\nabla_a+s A_a) K^{ab} (\nabla_b+s A_b)
 +2i N^{a}(\nabla_a+s A_a) - W \,,
\ee
where the second-derivative part is given by the Killing tensor (\ref{KillingtensoronKerr}), 
and $\bm{N}=N_adx^a$ and $W$ are given from Eq.~(\ref{tildeK_components}) as
\begin{align}
 \bm{N}
 &= \frac{2sp {\cal Q}}{r^2+p^2}(d\tau-p^2d\sigma)
 + \frac{2isr{\cal P}}{r^2+p^2}(d\tau+r^2d\sigma) \,, \\
 W &= -\frac{1}{r^2+p^2}\left(
\frac{2s^2+1}{6}(p^2{\cal Q}''-r^2{\cal P}'')
+\frac{2s^2rp({\cal P}'-i{\cal Q}')}{\chi}
-\frac{4s^2(p(p-2ir){\cal Q}+r(r+2ip){\cal P})}{\chi^2}
\right)
 \,.
\end{align}
In the present coordinates, the symmetry operator is explicitly given by
\begin{align}
 {\cal K}
 =& \frac{p^2}{r^2+p^2}\Bigg[{\cal Q}\partial_r^2 +(1+s){\cal Q}'\partial_r
  - \frac{1}{{\cal Q}}(r^2\partial_\tau -\partial_\sigma)^2
 + s \Bigg(\frac{{\cal Q}'}{{\cal Q}}\left(r^2\partial_\tau - \partial_\sigma\right)
   - 4 r\partial_\tau\Bigg)
  + \frac{(2s+1)(s+1)}{6} {\cal Q}''\Bigg]
\\
 & - \frac{r^2}{r^2+p^2}\Bigg[{\cal P}\partial_p^2+ {\cal P}'\partial_p
 + \frac{1}{{\cal P}}(p^2\partial_\tau +\partial_\sigma)^2 
 -i s \Bigg(
\frac{{\cal P}'}{{\cal P}}\left(p^2\partial_\tau + \partial_\sigma\right) 
- 4 p\partial_\tau
\Bigg)
   +\frac{(2s^2+1)}{6} {\cal P}'' 
   - s^2\frac{{\cal P}'^2}{4{\cal P}}\Bigg] 
\,.
   \nonumber
\end{align}

Direct calculations show that the Killing tensor (\ref{lifted-K_Teukolsky}) 
of the uplifted spacetime satisfies the anomaly-free condition (\ref{anomalyfreecond}), 
hence this operator commutes with the differential operator 
of the Teukolsky equation ${\cal H}$ defined by Eq.~(\ref{Teq}):
\be
 [{\cal H},{\cal K}] = 0 \,.
\ee

Using Eqs.\ (\ref{sepAnsatz}), (\ref{sepEquation_Kerr1}) and (\ref{sepEquation_Kerr2}),
we can confirm that the eigenvalue of the symmetry operator ${\cal K}$ coincides 
with the separation constant appearing in the Teukolsky equation
\be
 {\cal K}\Phi = \kappa \Phi \,.
\ee

\section{Symmetry operators for the LFKK equation in $D\geq 4$ dimensions}
\label{sec:sec4}

While the higher-dimensional generalization of the Teukolsky equation has not been found so far, 
the LFKK equation~(\ref{LFKKeq}) 
works for the $D$-dimensional Kerr-NUT-(A)dS spacetime with arbitrary dimensions $D$.
In this section,
we apply the Eisenhart-Duval lift summarized in Sec.~\ref{sec:sec2} to the LFKK in general dimensions.
As a result we obtain a covariant expression of the commuting symmetry operators of \cite{Krtous:2018bvk}, 
which played a key role to guarantee the separability of the LFKK equation.

After reviewing the metric of the $D$-dimensional Kerr-NUT-(A)dS spacetime in Sec.~\ref{sec:KNAdS-Ddim} 
and its hidden symmetry in Sec.~\ref{sec:HiddenSymofKNUTAdS}, 
we apply the Eisenhart-Duval lift to the LFKK equation defined on this spacetime
in Sec.~\ref{sec:lift-Ddim} to obtain the Killing tensors associated to the uplifted metric. 
Using them we can construct commuting symmetry operators, and in Sec.~\ref{sec:comparison} 
we confirm that they coincide with the symmetry operators of \cite{Krtous:2018bvk} 
up to the first-order differential operators, which is one of the main results of our work.
In Sec.~\ref{sec:Lorenzgauge} we briefly study the Lorenz gauge condition.
This section is supplemented with appendix~\ref{sec:anomaly-Ddim}, 
which is devoted to the analysis on the commutativity conditions 
that guarantee the differential operators obtained above become symmetry operators commuting with each other.

\subsection{Kerr-NUT-(A)dS metric in $D\geq 4$ dimensions}
\label{sec:KNAdS-Ddim}

The Kerr-NUT-(A)dS metric in $D=2n+\epsilon$ dimensions
is given by
\begin{equation}
\bm{g} = \sum_{\mu=1}^{n} \frac{d x_{\mu}^2}{Q_\mu(x)}+
\sum_{\mu=1}^{n} Q_{\mu}(x) \left( \sum_{k=0}^{n-1} \sigma^{(k)}_\mu
d \psi_k \right)^2+\epsilon S \left( \sum_{k=0}^{n} \sigma^{(k)} d \psi_k \right)^2,
\label{KNAdS_D}
\end{equation}
where $\epsilon=0$ for even dimensions and $\epsilon=1$ for odd dimensions, and
\begin{equation}
Q_\mu=\frac{X_\mu}{U_{\mu}},
\qquad
S=\frac{c}{\sigma^{(n)}},
\qquad
U_\mu=\prod_{\nu \ne \mu}(x_\mu^2-x_\nu^2)
\end{equation}
with a constant $c$ and arbitrary functions $X_\mu = X_\mu(x_\mu)$ depending only on $x_\mu$.
The functions $\sigma^{(k)}, \sigma^{(k)}_\mu$
are the $k$-th elementary symmetric functions defined as
\begin{equation}
\sigma^{(k)} =
\sum_{1\leq\nu_1<\nu_2<\cdots<\nu_k\leq n}
x_{\nu_1}^2 x_{\nu_2}^2\cdots x_{\nu_k}^2 \, ,
\qquad
\sigma^{(k)}_\mu
=
\sum_{
\substack{
1\leq\nu_1<\nu_2<\cdots<\nu_k\leq n \\ \nu_i\neq \mu
}
}
x_{\nu_1}^2 x_{\nu_2}^2\cdots x_{\nu_k}^2\,.
\end{equation}
To express the coordinates to express the metric~(\ref{KNAdS_D}), 
we reserve Latin indices $k,\ell,\ldots$ for the Killing directions $\psi_i$ 
and the Greek indices $\mu,\nu,\ldots$ for the non-Killing directions $x_\mu$.
We express the coordinates of the whole space with the Latin indices beginning $a$ as before, 
that is, $(x^a) = (x_\mu, \psi_k)$.

The inverse metric of (\ref{KNAdS_D}) is given by
\begin{equation}
\bm{g}^{-1}
= \sum_{\mu=1}^n g^{\mu \mu} \left( \frac{\partial}{\partial x_\mu} \right)^2
+ \sum_{k,\ell=0}^{n-1+\epsilon} g^{k \ell} \frac{\partial}{\partial \psi_k}\frac{\partial}{\partial \psi_{\ell}}
\,,
\label{inversemetricDdim}
\end{equation}
where
\begin{equation}
g^{\mu \mu}=Q_\mu,
\qquad
g^{k \ell}=\sum_{\mu=1}^{n-1+\epsilon} \zeta^{k \ell}_{(\mu)} Q_\mu,
\qquad
\zeta^{k \ell}_{(\mu)} = \frac{(-1)^{k+\ell}x_\mu^{2(2n-2-k-\ell)}}{X_\mu^2}
+\frac{(-1)^{n+1}}{c x_\mu^2 X_\mu} \delta_{nk} \delta_{n \ell} \,.
\end{equation}

To describe the metric (\ref{KNAdS_D}),
let us introduce the orthonormal basis defined as
\begin{equation}\label{basis1}
\bm{e}^{\underline\smu}
=\frac{dx^\mu}{\sqrt{Q_\mu}},
\qquad
\bm{e}^{\hat{\underline\smu}}
=\sqrt{Q_\mu}\left( \sum_{k=0}^{n-1} \sigma^{(k)}_\mu
d \psi_k \right),
\qquad
\bm{e}^{\underline 0}=\sqrt{S}\left( \sum_{k=0}^{n} \sigma^{(k)}
d \psi_k \right),
\end{equation}
where $\bm{e}^{\underline 0}$ exists only in odd dimensions ($\epsilon=1$).
This basis describes
the Euclideanized Kerr-NUT-(A)dS metric.
Their dual basis (vector field) is given by
\begin{equation}\label{basis2}
\bm{e}_{\underline\smu}
=\sqrt{Q_\mu} \frac{\partial}{\partial x_\mu},
\qquad
\bm{e}_{\hat{\underline\smu}}
=\sum_{k=0}^{n-1+\epsilon} \frac{(-1)^k x_\mu^{2(n-1-k)}}{\sqrt{Q_\mu} U_\mu}\frac{\partial}{\partial \psi_k},
\qquad
\bm{e}_{\underline 0}
=\frac{\sqrt{S}}{c} \frac{\partial}{\partial \psi_n}.
\end{equation}
Indices of the orthonormal basis and the tensor components 
with respect to it are marked with underlines 
to distinguish them from the coordinate basis indices $a, \mu, k$.
Using the orthonormal basis, the metric is expressed as
\begin{equation}
\bm{g} = \sum_{\mu=1}^n (\bm{e}^{\underline\smu} \bm{e}^{\underline\smu}
+ \bm{e}^{\hat{\underline\smu}} \bm{e}^{\hat{\underline\smu}})
+\epsilon \,\bm{e}^{\underline 0} \bm{e}^{\underline 0}\, .
\end{equation}

\subsection{Hidden symmetries of the Kerr-NUT-(A)dS spacetime}
\label{sec:HiddenSymofKNUTAdS}

The Kerr-NUT-(A)dS spacetime has hidden symmetries associated to the principal tensor~(\ref{principaltensor}).
See e.g.\ \cite{Frolov:2008jr,Frolov:2017kze,Yasui:2011pr} for details.
The principal tensor $\bm{h}=(1/2)h_{\underline{ab}}\bm{e}^{\underline a}\wedge \bm{e}^{\underline b}$ 
for the metric (\ref{KNAdS_D}) is written as
\begin{equation}
\bm{h} =\sum_{\mu=1}^n x_\mu \bm{e}^{\underline\smu} \wedge \bm{e}^{\hat{\underline\smu}}
\, .
\label{principaltensor-components}
\end{equation}
The associated 1-form
$\xi_a \equiv \frac{1}{D-1} \nabla^b h_{ba}$
introduced by Eq.~(\ref{principaltensor}) is
given by
\begin{equation}
\xi_{{\underline\smu}} = 0\,,
\qquad
\xi_{\hat {\underline\smu}} = \sqrt{Q_\mu}\,,
\qquad
\xi_{\underline 0} = \sqrt{S}
\,,
\end{equation}
which gives a Killing vector field
$\xi^a = g^{ab}\xi_b$.
We can check that the inverse metric (\ref{inversemetricDdim}) fits into Benenti's canonical form, 
and then the coordinate components of the commuting Killing tensors $K_{(j)}^{ab}$ $(j=0,1,\cdots, n-1)$ are obtained
as
\begin{equation}\label{Bkilling}
K_{(j)}^{\mu \mu}=\sigma^{(j)}_\mu Q_\mu,
\qquad
K_{(j)}^{k \ell}=\sum_{\mu=1}^{n-1+\epsilon} \zeta^{k \ell}_{(\mu)} \sigma^{(j)}_\mu Q_\mu
\, .
\end{equation}
If we use the orthonormal basis, they are written as
\begin{equation}\label{killing}
\bm{K}^{(j)}=\sum_{\mu=1}^n
\sigma^{(j)}_\mu
(\bm{e}^{\underline\smu} \bm{e}^{\underline\smu}
+ \bm{e}^{\hat{\underline\smu}} \bm{e}^{\hat{\underline\smu}})
+\epsilon \,\sigma^{(j)} \bm{e}^{\underline 0} \bm{e}^{\underline 0}~.
\end{equation}
It can be seen that $\eta_{(j)}^a$ defined as
the contraction of $\xi_a$ and $K_{(j)}^{ab}$,
\begin{equation}
\eta_{(j)}^a
\frac{\partial}{\partial x^a}
\equiv
\xi_a K_{(j)}^{ab} 
\frac{\partial}{\partial x^b}
=
\frac{\partial}{\partial \psi_j}
~,
\label{xiK_LFKK-Ddim}
\end{equation}
become Killing vectors.

\subsection{Eisenhart-Duval lift of Maxwell's equations 
on $D$-dimensional Kerr-NUT-(A)dS spacetime}
\label{sec:lift-Ddim}

In this section, we construct the uplifted spacetime 
based on the LFKK equation and construct the Killing tensors corresponding to it, 
which give the candidates of the symmetry operators for the LFKK equation.

The polarization tensor $\bm{B}$ is defined by Eq.~(\ref{B}).
Using 
Eq.~(\ref{principaltensor-components}),
the polarization tensor is calculated as
\begin{equation}
\label{pol2}
\bm{B} =
\frac{1}{1+\beta^2 x_\mu^2}\left(
  \bm{e}_{\underline \mu} \bm{e}_{\underline \mu}
+ \bm{e}_{\underline {\hat \mu}} \bm{e}_{\underline {\hat \mu}}
\right)
+ \frac{\beta x_\mu}{1 + \beta^2 x_\mu^2}
\left(
  \bm{e}_{\underline \mu} \bm{e}_{\underline{\hat \mu}}
- \bm{e}_{\underline{\hat \mu}} \bm{e}_{\underline \mu} 
\right)
+ \epsilon \,\bm{e}_{\underline 0} \bm{e}_{\underline 0} \, .
\end{equation}
The corresponding ansatz of the Maxwell potential 1-form 
$\bm{{\cal A}}={\cal A}_{\underline a} \bm{e}^{\underline a}$ is given by
\begin{align}
{\cal A}_{\underline\smu}
&= \frac{1}{1+\beta^2 x_\mu^2} \nabla_{\underline\smu} \Phi
+\frac{\beta x_\mu}{1+\beta^2 x_\mu^2} \nabla_{\hat{\underline\smu}} \Phi \\
{\cal A}_{\hat{\underline\smu}}
&= -\frac{\beta x_\mu}{1+\beta^2 x_\mu^2} \nabla_{\underline\smu} \Phi
+\frac{1}{1+\beta^2 x_\mu^2} \nabla_{\hat{\underline\smu}} \Phi \\
{\cal A}_{\underline 0}
&= \nabla_{\underline 0} \Phi \, .
\end{align}
This corresponds to the magnetic polarization in \cite{Krtous:2018bvk}.%
\footnote{
For the electric polarization,
the ansatz for the Maxwell field is taken as ${\cal A}_a = h_{ab}\nabla^b \Phi$, and then
Maxwell's equation is reduced to the Klein-Gordon equation $\Box \Phi=0$ 
and the Lorenz condition becomes $\xi^a\nabla_a \Phi=0$ \cite{Krtous:2018bvk}
The commuting symmetry operators for this equation can be found in, e.g., \cite{Frolov:2017kze,Yasui:2011pr}.
}
With this ansatz, Maxwell's equation is reduced to the LFKK equation~(\ref{LFKKeq}), which may be expressed as
\begin{equation}
{\cal H}\Phi \equiv
\left[
g^{ab}\left(
\nabla_a - iq A_a
\right)\left(
\nabla_b - iq A_b
\right)
-2V
\right]\Phi = 0
\, ,
\label{LFKKeq_Ddim_mod}
\end{equation}
where
\begin{equation}
qA^a = i\beta \xi_b B^{ba}
  \label{AVdef_LFKK-Ddim}
\end{equation}
and Eq.~(\ref{F_LFKK-Ddim}) implies
$V = iq \nabla^aA_a + q^2 A^aA_a$.
Using (\ref{pol2}) and the orthonormal basis (\ref{basis2}),
we have
\begin{align}
q\bm{A} &=
\sum_{\mu=1}^n
\frac{-i \beta^2 x_\mu \sqrt{Q_\mu}}{1+\beta^2 x_\mu^2}
\,\bm{e}^{\underline \smu}
+
\sum_{\mu=1}^n
 \frac{i \beta \sqrt{Q_\mu}}{1+\beta^2 x_\mu^2} 
\, \bm{e}^{\hat{\underline \smu}}
+
i
\beta \sqrt{S}
\, \bm{e}^{\underline 0}~,
\label{A_LFKK}
\\
  V &=
  -\frac{\beta^2}{2}\left[
  \sum_{\mu=1}^n \frac{x_\mu}{U_\mu} \frac{d}{dx_\mu} \left(
  \frac{X_\mu}{1+\beta^2 x_\mu^2}
  \right) + \epsilon 
\,\sum_{\mu=1}^n
\frac{1}{U_\mu}
\left(
\frac{X_\mu}{1+\beta^2x_\mu^2}
+ \frac{(-1)^nc}{x_\mu^2}
\right)\right] \,.
\end{align}

From (\ref{LFKKeq_Ddim_mod}),
we find the explicit form of the uplifted metric
as
\begin{align}
\tg^{\mu\mu} &= g^{\mu\mu}=Q_\mu ,   \nonumber\\
\tg^{kl} &= g^{kl}
=\sum_{\mu=1}^{n-1+\epsilon}
g^{\mu\mu}
\zeta^{k \ell}_{(\mu)}  ,
   \nonumber\\
\tg^{\mu v}&=   \frac{i\beta^2 x_\mu}{1+\beta^2 x_\mu^2}g^{\mu\mu}, \nonumber\\
\tg^{k v}&=-i \beta \sum_{\mu=1}^n  \frac{(-1)^k x_\mu^{2(n-1-k)}}{X_\mu(1+\beta^2 x_\mu^2})g^{\mu\mu}
\qquad
(k=0,1,\cdots,n-1)
~,
\nonumber\\
\tg^{nv}
&=
i\beta^3 \sum_{\mu=1}^n  \frac{(-1)^n}{X_\mu (1+\beta^2x_\mu^2)}g^{\mu\mu}
,\nonumber\\
\tg^{vv}
&=-\beta^2 \sum_{\mu}  \left( \frac{1}{X_\mu}\frac{d}{d x_\mu} 
\left( \frac{x_\mu X_\mu}{1+\beta^2 x_\mu^2} \right)+ \epsilon \frac{x_\mu}{1+\beta^2 x_\mu^2} \right)g^{\mu\mu}
, \nonumber\\
\tg^{uv}&= ~1~ = ~\sum_{\mu=1}^n  \frac{x_\mu^{2(n-1)}}{X_\mu}g^{\mu\mu}.
\label{upliftedmetric_LFKK}
\end{align}
In the above expressions,
$\tilde g^{kv}$ can be expressed in an alternative form as
\begin{equation}
\tg^{k v}
=
\frac{(-1)^n i \beta}{2} \sum_{\mu=1}^n g^{\mu\mu}
\frac{1-\beta^2 x_{\mu}^2}{X_\mu(1+\beta^2 x_\mu^2)}
\beta^{2(1-n+k)}
\equiv
\check g^{k v}
~.
\label{gkv_LFKK_2ndform}
\end{equation}
This form of $\tilde g^{kv}$ plays an important role 
in comparison of our result with \cite{Krtous:2018bvk} as we discuss in Sec.~\ref{sec:comparison}.

The inverse metric $\tg^{AB}$ 
for (\ref{upliftedmetric_LFKK})
takes the form
\begin{equation}
\tg^{\mu \mu}=Q_\mu,
\qquad
\tg^{A B}=\sum_{\mu=1}^{n} \zeta^{A B}_\mu Q_\mu
\qquad
(A=B \ne \mu) ,
\end{equation}
where $\zeta^{A B}_\mu$ is a function of only one variable $x_\mu$.
This fits into Benenti's canonical form of the inverse metric,%
\footnote{The inverse metric~(\ref{upliftedmetric_LFKK}) has $\tilde{g}^{\mu v}\neq 0$, 
which must be zero in Benenti's canonical form of the inverse metric.
However this component can be set to zero by a coordinate transformation
\begin{equation}
dv \rightarrow dv - \sum_{\mu=1}^n  \frac{i\beta^2 x_\mu}{1+\beta^2 x_\mu^2} dx_\mu~.
\notag
\end{equation}
This transformation does not affect the other metric components, 
hence the metric after this transformation manifestly fits into Benenti's canonical form, and then
the separability of geodesic equations for the metric is guaranteed.}
and then 
the mutually commuting Killing tensors $\tilde K^{AB}_{(j)}$ ~$(j=0,1, \cdots,n-1)$ are constructed as%
\footnote{In terms of the St\"ackel matrix for metric (\ref{KNAdS_D}), 
Eq.~(\ref{tildeK_LFKK}) is given by
\[
\tilde K^{\mu\mu}_{(j)} = \bar\phi^\mu{}_{(j+1)}\,,
\qquad
\tilde K^{A B}_{(j)}=\sum_{\mu=1}^{n} \zeta^{A B}_\mu (x_\mu)
\bar\phi^\mu{}_{(j+1)}~.
\]}
\begin{equation}
\tilde K^{\mu \mu}_{(j)}=\sigma^{(j)}_\mu Q_\mu \, ,
\qquad
\tilde K^{A B}_{(j)}=\sum_{\mu=1}^{n} \zeta^{A B}_\mu \sigma^{(j)}_\mu Q_\mu
\qquad
\Bigl(
(A,B) \neq (\mu,\mu)
\Bigr)
\,,
\label{tildeK_LFKK}
\end{equation}
where its components are given by
\begin{align}
\tilde K_{(j)}^{\mu\mu} &=
\sigma_\mu^{(j)}
Q_\mu ,   \nonumber\\
\tilde K_{(j)}^{kl} &=
\sum_{\mu=1}^{n-1+\epsilon}
\tilde K_{(j)}^{\mu\mu}
\zeta^{k \ell}_{(\mu)}
  ,
   \nonumber\\
\tilde K_{(j)}^{\mu v}&= i  \frac{\beta^2 x_\mu}{1+\beta^2 x_\mu^2}\tilde K_{(j)}^{\mu\mu}
, \nonumber\\
\tilde K_{(j)}^{k v}&= -i \beta \sum_{\mu=1}^n  \frac{(-1)^k x_\mu^{2(n-1-k)}}{X_\mu(1+\beta^2 x_\mu^2)}
\tilde K_{(j)}^{\mu\mu}
~,\nonumber\\
\tilde K_{(j)}^{nv}
&=
i\beta^3 \sum_{\mu=1}^n  \frac{(-1)^n}{X_\mu (1+\beta^2x_\mu^2)}\tilde K_{(j)}^{\mu\mu}
,\nonumber\\
\tilde K_{(j)}^{vv}
&= -\beta^2 \sum_{\mu}  \left( \frac{1}{X_\mu}\frac{d}{d x_\mu} 
\left( \frac{x_\mu X_\mu}{1+\beta^2 x_\mu^2} \right)+ \epsilon \frac{x_\mu}{1+\beta^2 x_\mu^2} \right)
\tilde K_{(j)}^{\mu\mu}
, \nonumber\\
\tilde K_{(j)}^{uv}&=
\delta_{j0}
=\sum_{\mu=1}^n \frac{x_\mu^{2(n-1)}}{X_\mu}\tilde K_{(j)}^{\mu\mu} .
\label{upliftedKT_LFKKDdim}
\end{align}
For $j=0$, the above expression reduces to
$\tilde K^{AB}_{(0)}=\tg^{AB}$.

Here,
we introduce the orthonormal basis of the uplifted spacetime by
\begin{equation}
\label{basis1_lift}
\tilde{\bm{e}}^{\underline a} = \bm{e}^{\underline a}\,, \qquad
\tilde{\bm{e}}^{+} =du \,, \qquad
\tilde{\bm{e}}^{-} = dv-V du+q A_a dx^a \, .
\end{equation}
with which the metric is given as
\be
 \bm{g} = \delta_{\underline a \underline b} \bm{e}^{\underline a} \bm{e}^{\underline b} \,, \qquad
\tilde{\bm{g}}
= \tilde{\eta}_{\underline A \underline B} \tilde{\bm{e}}^{\underline A}\tilde{\bm{e}}^{\underline B}
= \delta_{\underline a \underline b} 
\tilde{\bm{e}}^{\underline a} \tilde{\bm{e}}^{\underline b}
+ \tilde{\bm{e}}^+\tilde{\bm{e}}^- 
+ \tilde{\bm{e}}^-\tilde{\bm{e}}^+
\,.
\label{g_uplifted_orthonormal}
\ee
In this basis,
the Killing tensor (\ref{upliftedKT_LFKKDdim}) for the uplifted metric is expressed as
\begin{equation}
\tilde{\bm{K}}^{(j)}
\tilde{\bm{e}}^{\underline A} \tilde{\bm{e}}^{\underline B}
=
\sum_{\mu=1}^n \sigma^{(j)}_\mu(\tilde{\bm{e}}^{\underline\smu}  \tilde{\bm{e}}^{\underline\smu}
+\tilde{\bm{e}}^{\hat{{\underline\smu}}}  \tilde{\bm{e}}^{\hat{{\underline\smu}}})
+\epsilon \sigma^{(j)} \tilde{\bm{e}}^{\underline 0}  \tilde{\bm{e}}^{\underline 0}
+\delta_{j 0}(\tilde{\bm{e}}^{+}  \tilde{\bm{e}}^{-}
+\tilde{\bm{e}}^{-}  \tilde{\bm{e}}^{+})
+(1-\delta_{j0}) \tilde{K}^{(j)}_{++} \tilde{\bm{e}}^{+}  \tilde{\bm{e}}^{+},
\label{K_uplifted_orthonormal}
\end{equation}
where
\begin{eqnarray}
\tilde{K}^{(j)}_{++}
=
-\beta^2 \sum_{\mu}^n \frac{ x_\mu \sigma^{(j)}_\mu}{U_\mu} \frac{d}{d x_\mu} \left( \frac{X_\mu}{1+\beta^2 x_\mu^2} \right)
-\epsilon \beta^2 \sum_{\mu=1}^n \frac{ \sigma^{(j)}_\mu}{U_\mu}\left( \frac{X_\mu}{1+\beta^2 x_\mu^2}+\frac{(-1)^n c}{x_\mu^2} \right).
\label{Killing_LFKK-Ddim-lift}
\end{eqnarray}

Using the Killing tensors $\tilde{\bm{K}}^{(j)}$ we can construct symmetry operators ${\cal K}^{(j)}$ on the base space
following the procedure summarized in Sec.~\ref{sec:sec2}.
For the Killing tensors $\bm{K}^{(j)}$, gauge field $\bm{A}$ and scalar functions $W_{(j)}$ on the base space, 
${\cal K}^{(j)}$ take the form
\begin{equation}
{\cal K}^{(j)}
= (\nabla_a-iq A_a) K^{ab}_{(j)}(\nabla_b-iq A_b)-W_{(j)} \, ,
\label{sop_LFKK-Ddim-base}
\end{equation}
where $A^a$ is defined in (\ref{A_LFKK}) and
\begin{equation}
W_{(j)}=\hat{K}^{(j)}_{++}
\,.
\end{equation}
These operators are given by the general formula~(\ref{symop_explicit}), where $N^a_{(j)}$ identically vanishes in this case.

As explained in Sec.~\ref{sec:sec2}, 
the differential operators constructed from the Killing tensor~(\ref{upliftedKT_LFKKDdim}) commute with each other.
Hence, the operators ${\cal K}^{(j)}$ becomes symmetry operators commuting 
with the wave operator ${\cal H} = {\cal K}^{(0)}$ of the LFKK equation (\ref{LFKKeq_Ddim_mod}).
See appendix C.

Equation (\ref{sop_LFKK-Ddim-base}) may be expressed as
\begin{equation}
{\cal K}^{(j)}
=\nabla_a K^{ab}_{(j)}\nabla_b-2iq A_a K^{ab}_{(j)}\nabla_b+F_{(j)},
\label{calK_LFKKDdim}
\end{equation}
where $F_{(j)}$ are functions defined by
\begin{equation}
F_{(j)}=
- iq \nabla_a(K^{ab}_{(j)}A_b) - q^2 K^{ab}_{(j)} A_a A_b - W_{(j)}.
\end{equation}
Then direct calculations yield $F_{(j)}=0$, hence
\begin{align}
{\cal K}^{(j)}
&=\nabla_a K^{ab}_{(j)}\nabla_b
-2iq A_a
K^{ab}_{(j)} \nabla_b
\notag \\
&=\nabla_a K^{ab}_{(j)}\nabla_b
+
2\beta \xi_c B^c{}_a
K^{ab}_{(j)} \nabla_b
\notag \\
&=
\nabla_a K^{ab}_{(j)}\nabla_b
+
2\beta \xi_c 
K^{ca}_{(j)} 
B_a{}^b
\nabla_b
\notag \\
&=
\nabla_a K^{ab}_{(j)}\nabla_b
+
2\beta 
\eta^a_{(j)}
B_a{}^b
\nabla_b
~.
\label{calKj_LFKK-Ddim}
\end{align}
This gives the covariant expression of the commuting symmetry operators.
In (\ref{calKj_LFKK-Ddim}), we used Eq.~(\ref{AVdef_LFKK-Ddim}) at the second equality.
At the third equality of this equation, we have used the fact that $\bm{K}^{(j)}$ commute with $\bm{B}$, 
as we can see from their structures (\ref{killing}) and (\ref{pol2}).
The last equality follows from Eq.~(\ref{xiK_LFKK-Ddim}).

\subsection{Comparison with previous results}
\label{sec:comparison}

In the previous section we have constructed the symmetry operators 
for the LFKK equation that commute with each other. 
We compare these operators to those proposed in \cite{Krtous:2018bvk}.

We can express
the symmetry operators 
(\ref{calKj_LFKK-Ddim}) in terms of the coordinates as
\begin{equation}
{\cal K}^{(j)}
= \sum_{\mu=1}^n \frac{\sigma_j(\hat{x}_\mu)}{U_\mu} {\cal S}_\mu
~,
\label{sop-expression_LFKK-Ddim_0}
\end{equation}
where
\begin{align}
{\cal S}_\mu
&= (1+\beta^2 x_\mu^2) \frac{\partial}{\partial x_\mu} \left( \frac{X_\mu}{1+\beta^2 x_\mu^2}\frac{\partial}{\partial x_\mu}\right)
+\frac{1}{X_\mu} \left( \sum_{k=0}^{n-1+\epsilon} (-1)^k x_\mu^{2(n-1-k)} \frac{\partial}{\partial \psi_k} \right)^2 \nonumber\\
&\quad
 +\frac{2 \beta}{1+\beta^2 x_\mu^2} \sum_{k=0}^{n-1} (-1)^k x_\mu^{2(n-1-k)} \frac{\partial}{\partial \psi_k} + \epsilon
\left[
\frac{X_\mu}{x_\mu} \frac{\partial}{\partial x_\mu}-\frac{(-1)^n}{c x_\mu^2} \left( \frac{\partial}{\partial \psi_n} \right)^2
- 2 \frac{(-1)^n \beta^3}{1+\beta^2 x_\mu^2}\frac{\partial}{\partial \psi_n}
\right]
\, ,
\label{sop-expression_LFKK-Ddim}
\end{align}
where
the last term proportional to $\epsilon$
appears only in the odd-dimensional case.

In the previous subsection 
we mentioned that $\tilde g^{kv}$ in Eq.~(\ref{upliftedmetric_LFKK}) can be expressed in an alternative form $\check g^{kv}$ defined by Eq.~(\ref{gkv_LFKK_2ndform}).
Applying our procedure to $\check g^{kv}$,
we find the $kv$ component of the Killing tensor to become
\begin{equation}
\check{K}^{kv}_{(j)} \equiv
\frac{(-1)^n i \beta}{2} \sum_{\mu=1}^n \tilde{K}^{\mu \mu}_{(j)} \frac{1-\beta^2 x_\mu^2 }{X_\mu (1+\beta^2 x_\mu^2)}\beta^{2(1-n+k)}
~.
\label{checkK}
\end{equation}
We can show that $\check{K}^{kv}_{(j)}$ (Eq.~(\ref{checkK})) and $\tilde{K}^{kv}_{(j)}$ in Eq.~(\ref{upliftedKT_LFKKDdim}) differs by a constant unless $j=0$.\footnote{When $j=0$, $\tilde{K}^{kv}_{(j)}$ and $\check{K}^{kv}_{(j)}$ both reduce to the metric $\tilde g^{kv}$.}
For example, 
in four dimensions ($n=2$)
the difference is given by
\begin{equation}
\tilde{K}^{0v}_{(1)}-\check{K}^{0v}_{(1)}=\frac{i}{2 \beta}~,
\qquad
\tilde{K}^{1v}_{(1)}-\check{K}^{1v}_{(1)}=-\frac{i \beta}{2}~.
\end{equation}
Expression of the difference for general $n$ is not illuminating and we omit it here.
The symmetry operators can be constructed using $\check{K}^{kv}_{(j)}$, and its expression is given by Eq.~(\ref{sop-expression_LFKK-Ddim_0})
with ${\cal S}_\mu$ replaced by
\begin{align}
\check{\cal S}_\mu
&= (1+\beta^2 x_\mu^2) \frac{\partial}{\partial x_\mu} \left( \frac{X_\mu}{1+\beta^2 x_\mu^2}\frac{\partial}{\partial x_\mu}\right)
+\frac{1}{X_\mu} \left( \sum_{k=0}^{n-1+\epsilon} (-1)^k x_\mu^{2(n-1-k)} \frac{\partial}{\partial \psi_k} \right)^2 \nonumber\\
&\quad
-(-1)^n \beta \frac{1-\beta^2 x_\mu^2}{1+\beta^2 x_\mu^2}
 \sum_{k=0}^{n-1} 
\beta^{-2(n-1-k)}
\frac{\partial}{\partial \psi_k}
+ \epsilon
\left[
\frac{X_\mu}{x_\mu} \frac{\partial}{\partial x_\mu}-\frac{(-1)^n}{c x_\mu^2} \left( \frac{\partial}{\partial \psi_n} \right)^2
- 2 \frac{(-1)^n \beta^3}{1+\beta^2 x_\mu^2}\frac{\partial}{\partial \psi_n}
\right]
\, .
\label{sop-expression_LFKK-Ddim_2ndform}
\end{align}
In the even-dimensional case ($\epsilon=0$),
$\check{\cal S}_\mu$
coincides with the operators $\tilde {\cal C}_\mu$ of \cite{Krtous:2018bvk}, in which expressions in terms of the coordinates are given.

The difference between 
${\cal S}_\mu$
and 
$\check{\cal S}_\mu$
appears only in the third term on the right-hand side.
Due to this difference,
the symmetry operators constructed from 
${\cal S}_\mu$
differ from those constructed from
$\check {\cal S}_\mu$
by a linear combination with constant coefficients of first-order differential operators corresponding to the Killing vectors (${\cal L}^{(i)}$ in Sec.~\ref{sec:sec2}).
Hence the symmetry operators of \cite{Krtous:2018bvk} are given by Eq.~(\ref{calKj_LFKK-Ddim}) up to lower-order symmetry operators ${\cal L}^{(i)}$.

\subsection{Lorenz gauge condition}
\label{sec:Lorenzgauge}

Before closing this section we briefly
examine the Lorenz condition $\nabla_a {\cal A}^a = 0$, which must be imposed to obtain the LFKK equation~(\ref{LFKKeq}).
Using the ansatz~(\ref{Aansatz}) and also the symmetry operator 
(\ref{sop-expression_LFKK-Ddim_0}),
$\nabla_a {\cal A}^a = 0$ can be written as
\begin{equation}
\sum_{j=0}^{n-1} \beta^{2j} \left(
{\cal K}^{(j)}
Z+\beta \bigl(2(n-j)-3+\epsilon\bigr)\frac{\partial Z}{\partial \psi_j} \right)
+\epsilon \frac{\beta^{2n}}{c} \frac{\partial^2 Z}{\partial \psi_n{}^2} 
=0
\, .
\label{LorenzCondtion-LFKK-Ddim}
\end{equation}
In even dimensions,
this expression corresponds to ${\cal C}Z=0$ of \cite{Krtous:2018bvk} (up to contributions from lower-order symmetry operators mentioned in the previous subsection).

Using (\ref{separationconstants}), 
the symmetry operators in 
Eq.~(\ref{LorenzCondtion-LFKK-Ddim}) can be replaced with the separation constants as
\begin{equation}
\sum_{j=0}^{n-1} \beta^{2j} \left(
\kappa^{(j)}
+\beta \bigl(2(n-j)-3+\epsilon\bigr)
i\omega_j
\right)
-\epsilon \frac{\beta^{2n}}{c} 
\omega_j^2
=0~,
\end{equation}
which can be solved as an algebraic equation for $\beta$.
Since $\kappa^{(0)}=0$ because of the equation of motion ${\cal K}^{(0)}Z=0$, one root is given by $\beta=0$ and there are the other $D-2$ nontrivial roots.

\section{Summary}
\label{sec:summary}

In this work, we focused on the commuting symmetry operators 
that enabled the separation of variables in the LFKK formulation of the Maxwell perturbations 
on the Kerr-NUT-(A)dS spacetimes, and proposed 
a method to reproduce those operators 
in terms of the covariant quantities associated to the background geometry.
Applying the Eisenhart-Duval lift to the background metric, 
the perturbation equations can be reduced to the massless Klein-Gordon equations. 
Then the symmetry operators that commute with the d'Alembertian operator can be constructed 
from the Killing tensor associated to the uplifted metric, 
if they satisfy the commutativity conditions (\ref{anomalyfreecond}) and (\ref{commutativity}).

We showed that this procedure actually works for the Teukolsky equation in four dimensions, 
and also for the LFKK equation in the general dimensions.
For both of them, we found that the uplifted spacetime admits the Killing tensors, 
and also that they satisfy the commutativity conditions 
so that the differential operators constructed from them become commuting symmetry operators. 
We also confirmed that they coincide with those in \cite{Krtous:2018bvk} 
up to lower-order symmetry operators generated from the Killing vectors.
In this sense we found a geometric interpretation of the commuting symmetry operators 
proposed by \cite{Krtous:2018bvk}.

Formulation in Sec.~\ref{sec:sec2} can be applied to more general equations, not only to the Maxwell equation.
Hence it may be possible to construct symmetry operators in a more transparent manner for various equations of motion.
For example, 
in four dimensions the perturbation equations of spin-$s$ fields with arbitrary $s$ can be reduced to the Teukolsky equations.
Then the uplifted metrics and the symmetry operators can be constructed as we did for the Maxwell (spin-1) field.
Particularly the gravitational perturbations ($s=2$) can be studied in this manner. 
%
It would be interesting to examine if we could learn how to generalize the LFKK formulation 
to the gravitational perturbations of rotating black holes in four and higher dimensions,
which is 
yet to be clarified.

\section*{Acknowledgments}

This work was partly supported by Osaka City University Advanced Mathematical Institute
 (MEXT Joint Usage/Research Center on Mathematics and Theoretical Physics).
N.T. is supported by Grant-in-Aid for Scientific Research from the Ministry of Education, 
Culture, Sports, Science and Technology, Japan No.18K03623
and AY 2018 Qdai-jump Research Program of Kyushu University.
Y.Y. is supported by Grant-in-Aid for Scientific Research from the Ministry of Education, 
Culture, Sports, Science and Technology, Japan No.16K05332.


\appendix

\section{Killing equations on the uplifted spacetime}
\label{app:Killingeq}

We investigate the contravariant components of the Killing equations 
for a Killing vector $\tilde{L}^A$ and a Killing tensor $\hK_{AB}$
on the $(D+2)$-dimensional uplifted spacetime $(\tilde{M},\tilde{\bm{g}})$
with the metric (\ref{liftedmetric}),
\be
 \tilde{\nabla}^{(A}\tilde{L}^{B)} = 0 \, ,
\qquad
 \tilde{\nabla}^{(A}\tilde{K}^{BC)} = 0 \,.
 \label{KE}
\ee
Assuming that the components of $\tilde{L}^A$ and $\tilde{K}^{AB}$ 
do not depend on the coordinates $u$ and $v$ [cf.~(\ref{compindepuv})], 
we can recast the components of $\tilde{L}^A$ and $\tilde{K}^{AB}$ 
by the variables on the base space $(M,\bm{g})$ introduced by (\ref{killing1_mod}),
$(L^a,I,J)$ and $(K^{ab},U^a,N^a,C,T,W)$.
Then the Killing equation (\ref{KE}) for the Killing vector $\tilde L^A$ is summarized as
\begin{align}
 \nabla_{(a}L_{b)} &= 0\, , \label{KL01}\\
\partial_b I &= 0\, ,  \label{KL02}\\
\partial_a J - q F_{ab}L^b  &= 0 \,,  \label{KL03}\\
L^a \partial_a V &=0\,. \label{KL04}
\end{align}
These show that 
$L^a$ is a Killing vector on the base space, 
$I$ is a constant and $V$ is constant along the Killing vector $L^a$.
Similarly, the Killing equation (\ref{KE}) 
for the Killing tensor $\tilde K_{AB}$ is summarized as
\begin{align}
 \nabla_{(a}K_{bc)}&=0 \,, \label{KE01} \\
 \nabla_{(a}N_{b)} + U_{(a}\partial_{b)}V 
 + q F^c{}_{(a}K_{b)c} &= 0 \,, \label{KE02} \\
 \nabla_{(a}U_{b)} &=0 \,, \label{KE03} \\
 \partial_aT + 2C\partial_aV -q F_{ab}U^b &= 0 \,, \label{KE04} \\
 \partial_aW -2(K_{ab}-CVg_{ab})\partial^bV
 -2q F_{ab}(N^b + VU^b)&=0\,, \label{KE05} \\
 \partial_a C &=0 \,, \label{KE06} \\
 U^a\partial_a V &= 0 \,, \label{KE07} \\
 N^a\partial_a V &= 0 \label{KE08} \,.
\end{align}
These show that
$K_{ab}$ is a Killing tensor,
$U^a$ is a Killing vector,
$C$ is a constant and
$V$ is constant along the vectors $U^a$ and $N^a$.
Moreover, taking the trace of Eq.\ (\ref{KE02}) and then using Eq.\ (\ref{KE07}),
 we obtain
\be
 \nabla_aN^a = 0 \,. \label{divNzero}
\ee
This divergence-fee condition for the vector $N^a$ becomes important to obtain the
expression (\ref{symop_explicit}) for the differential operator ${\cal K}$.

\section{Teukolsky equations for Maxwell fields}
\label{App:Teukolsky}

Benn, Charlton and Kress discussed the Hertz equation
on a spacetime admitting a gauged conformal Killing-Yano tensor (GCKY),
instead of treating the Maxwell equation.
They demonstrated in \cite{Benn:1996su} that, 
if a spacetime in four dimensions admits 
such a tensor with certain additional conditions,
the Hertz equation reduces to a scalar-type equation called Debye equation.
Since the Kerr spacetime admits three GCKYs,
we obtain three Debye equations.
It is interesting that two of them can be solved by separation of variables,
which are known as the Teukolsky equations for Maxwell fields~\cite{PhysRevLett.29.1114,Teukolsky:1973ha}.
We review its derivation following \cite{Benn:1996su}.

\subsection{Reduction of the Hertz equation with GCKY}

The Hertz equation is given by
\be
 \Delta \bm{{\cal P}} = d \bm{{\cal G}} + \delta \bm{{\cal W}} \,,
 \label{HertzEquation}
\ee
where $\bm{{\cal P}}$, $\bm{{\cal G}}$ and $\bm{{\cal W}}$ are
2-form, 1-form and 3-form, and $\bm{{\cal P }}$ is called Hertz potential.
Here, $d$ and $\delta$ are the exterior derivative and co-derivative,
i.e., for a $p$-form $\bm{\alpha}$,
\begin{align}
 (d\bm{\alpha})_{a_1\dots a_{p+1}} &= (p+1)\nabla_{[a_1}\alpha_{a_2\dots a_{p+1}]} \,, \\
 (\delta \bm{\alpha})_{a_1\dots a_{p-1}} &= - \nabla^b\alpha_{ba_1\dots a_{p-1}} \,,
\end{align}
and $\Delta$ is the Laplace-de Rham operator,
i.e., $\Delta\equiv -d\delta -\delta d$.

Given a Hertz potential $\bm{{\cal P}}$,
one can construct the gauge potential for a Maxwell field by
\be
 \bm{{\cal A}} = \delta \bm{{\cal P}} + \bm{{\cal G}} \,.
 \label{AfromPandG}
\ee
Since its field strength $\bm{{\cal F}}\equiv d\bm{{\cal A}}$ is given by
\be
 \bm{{\cal F}} = d(\delta \bm{{\cal P}} + \bm{{\cal G}})
 = -\delta( d \bm{{\cal P}} + \bm{{\cal W}}) \,,
\ee
this $\bm{{\cal F}}$ solves the Maxwell equation,
\be
 d\bm{{\cal F}} = 0 \,, \qquad \delta \bm{{\cal F}} = 0 \,.
 \label{MaxwellEquations}
\ee
Hence, we may solve the Hertz equation 
for $\bm{{\cal P}}$,
instead of solving the Maxwell equation for $\bm{{\cal F}}$.

To solve the Hertz equation (\ref{HertzEquation}), 
Benn, Charlton and Kress \cite{Benn:1996su} considered
 a rank-2 GCKY $\bm{h}$ to make an ansatz 
for the Hertz potential $\bm{{\cal P}}$.
A rank-2 GCKY $\bm{h}$ is the 2-form satisfying the equation
\begin{equation}
\hat\nabla_{(a}h_{b)c} = 
\xi_c g_{ab} - \xi_{(a}g_{b)c} ~,
\qquad
\xi_a = \frac13 \hat\nabla^b h_{ba} ~,
 \label{CKYequation_gauge}
\end{equation}
where the hatted nabla denotes the gauge-covariant derivative
 $\hat{\nabla} \equiv \nabla + \bm{A}$.
Note that, since the gauge field in the gauge-covariant derivative and the gauge potential
of Maxwell field are different objects in general,
we denote the former by $\bm{A}$ and the latter by $\bm{{\cal A}}$ in order to distinguish them.

Now, suppose that a $4$-dimensional spacetime $(M,\bm{g})$ admits
a rank-2 GCKY $\bm{h}$
and it satisfies the eigenvalue equation
\be
 {\cal M} \bm{h} = \lambda_h \bm{h}
 \label{eigenvalueequation}
\ee
with some function $\lambda_h$, where ${\cal M}$ is a linear map
on the space of 2-forms given by
\begin{equation}
 ({\cal M}\bm{h})_{ab} =
\frac{1}{2} 
C_{ab}{}^{cd} h_{cd}
-\frac16R\,h_{ab} 
-  F^c{}_{[a}h_{b]c} \,.
\end{equation}
Here, $C_{abcd}$
and $R$ are the Weyl, Ricci and scalar curvatures
and $\bm{F}=d\bm{A}$.
For such a rank-2 GCKY $\bm{h}$ and a scalar $\Phi$,
we obtain
\be
 \Delta (\Phi \bm{h}) = \left(\check{\Box}\Phi 
 + \lambda_h\Phi \right) \bm{h} + d\bm{G} + \delta\bm{W} 
\,,
 \label{pre15}
\ee
where
\begin{equation}
G_a = 2\Phi \xi_a\,,
\qquad
W_{abc} = -2\Phi \hat\nabla_{[a}h_{bc]}\,.
 \label{Def_WandG_gauge}
\end{equation}
Here, $\check{\Box}$ is the Laplace-Beltrami operator with respect to the gauge-covariant derivative
$\check \nabla \equiv \nabla - \bm{A}$, i.e. $\check{\Box}=g^{ab}
\check\nabla_a \check \nabla_b$.
Hence, it turns out that $\bm{{\cal P}} = \Phi \bm{h}$ becomes
the Hertz potential satisfying Eq.\ (\ref{HertzEquation}) 
with $\bm{{\cal G}}=\bm{G}$
and $\bm{{\cal W}}=\bm{W}$
if and only if the scalar $\Phi$ satisfies the equation
\be
\check{\Box}\Phi + \lambda_h \Phi = 0\,.
\label{HertzEquation2}
\ee
This equation is called the Debye equation. 
When we consider the Maxwell equation in the Kerr spacetime, 
the Debye equation corresponds to the Teukolsky equation for $s=\pm 1$.

It is important to notice that the GCKY equation (\ref{CKYequation_gauge})
is invariant under the gauge transformation
\be
 \bm{h} \to \bm{h}^\prime = \Omega \bm{h} \,, \qquad
 \bm{A} \to \bm{A}^\prime = \bm{A} - d \log \Omega
\ee
for an arbitrary function $\Omega$. 
After this gauge transformation, $\bm{h}^\prime$ and $\bm{A}^\prime$ 
still satisfy the condition (\ref{eigenvalueequation})
with the same $\lambda_h$ and hence
the Debye equation (\ref{HertzEquation2}) becomes
\be
\check{\Box}^\prime \Phi^\prime + \lambda_h \Phi^\prime = 0 \,,
\label{HertzEquation2_gaugetrsf}
\ee
with $\Phi^\prime = \Omega^{-1}\Phi$,
where $\check{\Box}^\prime$ is the Laplace-Beltrami operator
of the gauge-covariant derivative $\check{\nabla}^\prime = \nabla - \bm{A}^\prime$.

In \cite{Benn:1996su}, it was shown that if a repeated principal null direction (RPND) is shear free
and geodesic, one can construct a GCKY from such a RPND 
and it satisfies the eigenvalue equation (\ref{eigenvalueequation}).
Hence, there might be at least one GCKY in algebraically special spacetimes
of type II, III, D and N.
Particularly, since type D spacetimes have two RPNDs,
one might have two GCKYs.
Moreover, it was shown that given two GCKYs,
one can construct a third GCKY from the two GCKYs.
Remembering the Goldberg-Sachs theorem, 
the RPNDs in a Ricci-flat type D spacetime are shear free and geodesic,
so one has three GCKYs.
In the present paper, we consider the off-shell Kerr-NUT-(A)dS spacetime
with the Carter metric,
which is known as the type D spacetime admitting a rank-2 Killing-Yano tensor.
As shown below, the RPNDs in this spacetime are shear free and geodesic,
so we obtain three GCKYs.

\subsection{Null tetrad and its spin coefficients for the Carter metric}

The null tetrad vectors $\{\bm{k},\bm{l},\bm{m},\bar{\bm{m}}\}$%
\footnote{The null tetrad is same as the one used in \cite{Chandrasekhar:1985kt} (see eq.\ (170) in page 299).
In \cite{Chandrasekhar:1985kt}, however, the inner products between the basis are chosen as
$g_{ab}k^a l^b = 1$ and $g_{ab}m^a \bar{m}^b = -1$ with the signature of the metric $(+,-,-,-)$.
Contrary to this, since we are now using the signature of $(-,+,+,+)$, we have (\ref{gklgmm}).} 
are given by
\begin{equation}
\bm{k}
= 
\sqrt\frac{r^2+p^2}{{\cal Q}(r)} \left(
\bm{e}_{\underline 0} + \bm{e}_{\underline 1}
\right)~,
\quad
\bm{l}
=
\frac12
\sqrt\frac{{\cal Q}(r)}{r^2+p^2} \left(
\bm{e}_{\underline 0} - \bm{e}_{\underline 1}
\right)~,
\quad
\bm{m}= 
\frac{\sqrt{r^2+p^2}}{\sqrt2 \bar \chi} \left(
\bm{e}_{\underline 2} - i \bm{e}_{\underline 3}
\right)~,
 \label{nulltetradCartervectors}
\end{equation}
where 
$\chi$ is defined in Eq.~(\ref{ATVT})
and $\bar{\bm{m}}$ is complex conjugate of $\bm{m}$.
The basis of (\ref{nulltetradCartervectors}) satisfy that
\begin{equation}
 g_{ab}k^a l^b = -1 \,, \qquad
 g_{ab}m^a \bar{m}^b = 1 \,, \label{gklgmm}
\end{equation}
and the other inner products are zero.
We introduce the null 1-forms
$\{\bm{k}_*,\bm{l}_*,\bm{m}_*,\bar{\bm{m}}_*\}$
dual to the basis of (\ref{nulltetradCartervectors})
in the sense that for the null vectors $\{\bm{E}_a\}=\{\bm{k},\bm{l},\bm{m},\bar{\bm{m}}\}$,
their dual 1-forms $\{\bm{E}^a_*\}=\{\bm{k}_*,\bm{l}_*,\bm{m}_*,\bar{\bm{m}}_*\}$
are given by $\bm{E}^a_*(\bm{E}_b)=\delta^a_b$ are
\begin{equation}
\bm{k}_*= 
\frac12 \sqrt\frac{{\cal Q}(r)}{r^2+p^2} 
\left(\bm{e}^{\underline 0} + \bm{e}^{\underline 1}\right)~,
\quad
\bm{l}_*= 
\sqrt\frac{r^2+p^2}{{\cal Q}(r)}
\left(\bm{e}^{\underline 0}- \bm{e}^{\underline 1}\right)~,
\quad
\bm{m}_*= \frac{\sqrt{r^2+p^2}}{\sqrt2 \chi}
\left(
\bm{e}^{\underline 2} + i \bm{e}^{\underline 3}
\right)~,
 \label{nulltetradCarter2}
\end{equation}
with which the metric is given by
\begin{equation}
 \bm{g}=-2\bm{k}_*\bm{l}_*
 +2\bm{m}_*\bar{\bm{m}}_*
  \,. 
\end{equation}

With this null tetrad, 
the Weyl scalars are zero except for
\begin{equation}
\Psi_2 = - C_{abcd}k^a m^b \bar m^c l^d \,.
\end{equation}
This means that $\bm{k}$ and $\bm{l}$ are RPNDs, 
so that the off-shell Kerr-NUT-(A)dS spacetime is of type D.

For this null tetrad, the corresponding spin coefficients\footnote{
The spin coefficients are given by
\begin{align}
&\kappa = g(\nabla_kk,m),\quad
 \sigma = g(\nabla_mk,m),\quad
 \lambda = g(\nabla_{\bar{m}}\bar{m},l),\quad
 \nu = g(\nabla_l\bar{m},l),\nonumber\\
&\rho = g(\nabla_{\bar{m}}k,m),\quad
 \mu = g(\nabla_m\bar{m},l),\quad
 \tau = g(\nabla_lk,m),\quad
 \pi = g(\nabla_k\bar{m},l),\nonumber\\
&\epsilon = \frac{1}{2}(g(\nabla_kk,l)+g(\nabla_k\bar{m},m)),\quad
 \gamma = \frac{1}{2}(g(\nabla_lk,l)+g(\nabla_l\bar{m},m)),\nonumber\\
&\alpha = \frac{1}{2}(g(\nabla_{\bar{m}}k,l)+g(\nabla_{\bar{m}}\bar{m},m)),\quad
 \beta = \frac{1}{2}(g(\nabla_mk,l)+g(\nabla_m\bar{m},m)) \,. \nonumber
\end{align}
} are given by
\begin{align}
&\kappa 
= \sigma
= \lambda
= \nu = 0 \,, \nonumber\\
& \alpha =-
\frac{\sqrt{r^2 + p^2}}{2 \sqrt2 \chi} \left( 
\frac{p+2ir}{r^2 + p^2} - \partial_p 
\right)\sqrt\frac{{\cal P}(p)}{r^2 + p^2}
~, \quad
\beta = -
\frac{\sqrt{r^2 + p^2}}{2 \sqrt2 \bar\chi} 
\left( \frac{p}{r^2 + p^2} + \partial_p \right)
\sqrt\frac{{\cal P}(p)}{r^2 + p^2}
~,
\notag \\
& \epsilon = 0\,, \quad
\gamma = 
\frac{{\cal Q}(r)}{2\chi^2 \bar\chi} - \frac{{\cal Q}'(r)}{4\left(r^2+p^2\right)}\,,
\quad
\rho = \frac1\chi\,,
\quad
\mu = \frac{{\cal Q}(r)}{2\chi^2\bar\chi}\,,
\notag \\
& \tau = 
\frac{i}{\sqrt2 \sqrt{r^2+p^2}} \sqrt\frac{{\cal P}(p)}{r^2 + p^2}
~,
\quad
\pi = -
\frac{i\bar\chi}{\sqrt2 \chi\sqrt{r^2+p^2}} \sqrt\frac{{\cal P}(p)}{r^2+p^2}\,.
\label{spincoefficients}
\end{align}
Since $\sigma=\lambda=0$, ${\bm k}$ and ${\bm l}$ are shear free.
Since $\kappa=\nu=0$, ${\bm k}$ and ${\bm l}$ are geodesic.
Hence, ${\bm k}$ and ${\bm l}$ are shear free and geodesic.

\subsection{GCKYs for the Carter metric and the Teukolsky equations}

The Carter metric (\ref{Cartermetric}) admits three GCKYs $\bm{h}^{(1)}$, 
$\bm{h}^{(2)}$ and $\bm{h}^{(3)}$, which are given by
\begin{align}
 \bm{h}^{(1)} &= \bm{l}_* \wedge \bar{\bm{m}}_* \,, \\
 \bm{h}^{(2)} &= \bm{k}_* \wedge \bm{m}_* \,, \\
 \bm{h}^{(3)} &= \bm{k}_* \wedge \bm{l}_* 
 - \bm{m}_* \wedge \bar{\bm{m}}_* \,,
\end{align}
with the gauge fields
\begin{align}
 \bm{A}^{(1)} &= 2(\epsilon + \rho)\bm{k}_* + 2 \gamma \bm{l}_* 
 + 2(\beta + \tau) \bm{m}_* + 2\alpha \bar{\bm{m}}_* \,, \label{gp_gcky_1}\\
 \bm{A}^{(2)} &=  -2\epsilon\bm{k}_* -2(\gamma+\mu) \bm{l}_*
  -2\beta  \bm{m}_* -2(\alpha+\pi)  \bar{\bm{m}}_* \,, \label{gp_gcky_2}\\
 \bm{A}^{(3)} &= \rho \bm{k}_* -\mu \bm{l}_*
  + \tau \bm{m}_* -\pi  \bar{\bm{m}}_* \,. \label{gp_gcky_3}
\end{align}
They satisfy 
\begin{equation}
\bm{A}^{(1)} + \bm{A}^{(2)} = 2\bm{A}^{(3)}~,
\end{equation}
and $\bm{A}^{(3)}$ is pure-gauge,
expressed as $\bm{A}^{(3)} = d\log \chi$.
This implies that their field strengths satisfy 
$\bm{F}^{(1)}=-\bm{F}^{(2)}$ and 
${\bm F}^{(3)} = 0$.
Since we can check that $\bm{h}^{(i)}$ for $i=1,2,3$ satisfy
\be
 {\cal M} \bm{h}^{(i)} = \lambda^{(i)} \bm{h}^{(i)}
\ee
with
\begin{equation}
\lambda^{(1,2)} = -\frac{R}{6}-4\Psi_2 ~, \qquad
\lambda^{(3)}= -\frac{R}{6}+2\Psi_2 ~,\label{ev_gcky}
\end{equation}
the corresponding Debye equations are given by [cf.~(\ref{HertzEquation2})]
\be
{\cal H}^{(i)}\Phi \equiv  
\left(\check\Box^{(i)} + \lambda^{(i)} \right)\Phi = 0 \,,
\ee
where $\check\Box^{(i)} \equiv g^{ab}\left(\nabla_a - A^{(i)}_a\right)\left(\nabla_b - A^{(i)}_b\right)$.
Especially, the Debye equation for $\bm{h}^{(1)}$ is explicitly given by
\begin{align}
 {\cal H}^{(1)}\Phi =&
 \frac{1}{r^2+p^2}\Bigg[{\cal Q}\partial_r^2 +{\cal P}\partial_p^2
 + {\cal P}'\partial_p
 - \frac{1}{{\cal Q}}(r^2\partial_\tau -\partial_\sigma)^2
 + \frac{1}{{\cal P}}(p^2\partial_\tau +\partial_\sigma)^2 \nonumber\\
 & -\frac{{\cal Q}'}{{\cal Q}}(r^2\partial_\tau -\partial_\sigma)
   +i  \frac{{\cal P}'}{{\cal P}}(p^2\partial_\tau+\partial_\sigma)
   + 4(r-ip)\partial_\tau
   + \frac{2{\cal P}{\cal P}''-{\cal P}'^2}{4{\cal P}} \Bigg]\Phi \,,
\end{align}
which can be solved by separation of variables.
This equation coincides with the Teukolsky equation for $s=-1$
when the Carter metric is restricted to the Kerr metric, i.e.,
${\cal Q}$ and ${\cal P}$ are given by (\ref{funcQandP}).

The Debye equations for $\bm{h}^{(2)}$ and $\bm{h}^{(3)}$
are not separating the variables.
To obtain the Teukolsky equation for $s=1$,
we need to perform the gauge transformation
\be
 \bm{h}^{(2)} \to \bm{h}^{(2)\prime} = \chi^2 \bm{h}^{(2)} \,, \qquad
 \bm{A}^{(2)} \to \bm{A}^{(2)\prime} = \bm{A}^{(2)} - 2\bm{A}^{(3)} \,,
\ee
and then obtain the GCKY $\bm{h}^{(2)\prime}$
with $\bm{A}^{(2)\prime}$ satisfying $\bm{A}^{(1)}+\bm{A}^{(2)\prime}=0$.
Here, $\bm{A}^{(2)\prime}=-\bm{A}^{(1)}$ is crucial 
for separating the variables in the Debye equation (see \cite{Benn:1996su} for details).
Actually, the Debye equation for $\bm{h}^{(2)\prime}$ with $\bm{A}^{(2)\prime}$
is explicitly given by
\begin{align}
 {\cal H}^{(2)\prime}\Phi =&
 \frac{1}{r^2+p^2}\Bigg[{\cal Q}\partial_r^2 +{\cal P}\partial_p^2
 +2{\cal Q}'\partial_r+ {\cal P}'\partial_p
 - \frac{1}{{\cal Q}}(r^2\partial_\tau -\partial_\sigma)^2
 + \frac{1}{{\cal P}}(p^2\partial_\tau +\partial_\sigma)^2 \nonumber\\
 & + \frac{{\cal Q}'}{{\cal Q}}(r^2\partial_\tau -\partial_\sigma)
   - i  \frac{{\cal P}'}{{\cal P}}(p^2\partial_\tau+\partial_\sigma)
   - 4(r-ip)\partial_\tau
   + \frac{2{\cal P}{\cal P}''-{\cal P}'^2}{4{\cal P}}
   + {\cal Q}'' \Bigg]\Phi \,.
\end{align}
When we consider the Kerr metric case in which
${\cal Q}$ and ${\cal P}$ are given by (\ref{funcQandP}), 
this coincides with the Teukolsky equation for $s=1$
and can be solved by separation of variables.

\section{Calculations on the commutativity conditions}
\label{sec:anomaly-Ddim}

In this appendix, we examine whether or not
the commutativity conditions (\ref{commutativity})  are satisfied 
with the Killing tensor (\ref{K_uplifted_orthonormal}) of the uplifted metric~(\ref{upliftedmetric_LFKK}).
For this purpose we introduce the orthonormal basis in Sec.~\ref{app:Curvature_lifted}, 
and then evaluate the commutativity conditions (\ref{commutativity}) by direct calculations in Sec.~\ref{app:commutativity}.

\subsection{Orthonormal basis, covariant derivatives, and curvature quantities
of the uplifted metric}
\label{app:Curvature_lifted}

We introduce the orthonormal basis of 1-forms $\{\tilde{\bm{e}}^{\underline A} \}
=\{\tilde{\bm{e}}^{\underline a},\tilde{\bm{e}}^+,\tilde{\bm{e}}^-\}$ by
Eq.~(\ref{basis1_lift}),
with which the base metric $\bm{g}$ and the uplifted metric $ \tilde{\bm{g}}$ are given by
Eq.~(\ref{g_uplifted_orthonormal}).
with $\eta_{\underline a \underline b}=\textrm{diag}(-1,1,\dots,1)$.
Hence, the dual basis $\{\tilde{\bm{e}}_{\underline A} \}
=\{\tilde{\bm{e}}_{\underline a},\tilde{\bm{e}}_+,\tilde{\bm{e}}_-\}$ are given by
\begin{equation}
\label{basis2_lift}
\tilde{\bm{e}}_{\underline a}
=\bm{e}_{\underline a}-q A_{\underline a} \partial_v \,, \qquad
\tilde{\bm{e}}_{+}=V \partial_v+\partial_u \,, \qquad
\tilde{\bm{e}}_{-}= \partial_v\,.
\end{equation}
With this basis,
the first structure equation $\tilde{d}\tilde{\bm{e}}^{\underline A} 
+ \tilde{\bm{\omega}}^{\underline A}{}_{\underline B} \wedge \tilde{\bm{e}}^{\underline B}= 0$
and $\tilde{\bm{\omega}}_{\underline B \underline A} = - \tilde{\bm{\omega}}_{\underline A \underline B}$ give us
the connection 1-forms on $(\tilde{M},\tilde{\bm{g}})$,
\begin{align}
\tilde{\bm{\omega}}_{\underline a \underline b}
= \bm{\omega}_{\underline a \underline b}-\frac{q}{2}F_{\underline a \underline b}\,\tilde{\bm{e}}^{+}\,, \qquad
\tilde{\bm{\omega}}_{+\underline a}
= 
-(\nabla_{\underline a} V) 
\,\tilde{\bm{e}}^{+}
+\frac{q}{2}F_{\underline a \underline b} \,\bm{e}^{\underline b}\,, \qquad
\tilde{\bm{\omega}}_{- \underline a}
= \tilde{\bm{\omega}}_{+-}=0\,,
\end{align}
where $\bm{\omega}_{\underline a \underline b}$ are the connection 1-forms on $(M,\bm{g})$
and $F_{\underline a \underline b}$ are the components of the field strength $\bm{F}=d\bm{A}$
on $(M,\bm{g})$.
Using the formula $\tilde{\nabla}_{\tilde{\bm{e}}_{\underline A}} \tilde{\bm{e}}^{\underline B}
= - \tilde{\bm{\omega}}^{\underline B}{}_{\underline C}(\tilde{\bm{e}}_{\underline A}) \,\tilde{\bm{e}}^{\underline C}$,
we have
\begin{align}\label{der1}
\tilde{\nabla}_{\tilde{\bm{e}}_{\underline a}} \tilde{\bm{e}}^{\underline b}
&=-\bm{\omega}^{\underline b}{}_{\underline c}(\bm{e}_{\underline a}) \,\tilde{\bm{e}}^{\underline c}
-\frac{q}{2}F_{\underline a}{}^{\underline b}\,\tilde{\bm{e}}^{+}~, \\
\tilde{\nabla}_{\tilde{\bm{e}}_{\underline a}} \tilde{\bm{e}}^{-}
&=\frac{q}{2}F_{\underline a \underline b} \,\tilde{\bm{e}}^{\underline b}~, \\
\tilde{\nabla}_{\tilde{\bm{e}}_+} \tilde{\bm{e}}^{\underline a}
&= \frac{q}{2}F^{\underline a}{}_{\underline b}\,\tilde{\bm{e}}^{\underline b}
-\delta^{\underline a \underline b}
(\nabla_{\underline b} V) 
\,\tilde{\bm{e}}^{+} ~, \\
\tilde{\nabla}_{\tilde{\bm{e}}_+} \tilde{\bm{e}}^{-}
&=
(\nabla_{\underline a} V) 
\,\tilde{\bm{e}}^{\underline a}~,
\end{align}
and the others are zero.

The curvature 2-forms $\tilde{\bm{R}}_{\underline A \underline B}
= \tilde{d}\tilde{\bm{\omega}}_{\underline A \underline B}
+ \tilde{\bm{\omega}}_{\underline A}{}^{\underline C} \wedge \tilde{\bm{\omega}}_{\underline C \underline B}$
on $(\tilde{M},\tilde{\bm{g}})$ are then given by
\begin{align}
& \tilde{\bm{R}}_{\underline a \underline b}
= \bm{R}_{\underline a \underline b}
-\frac{q}{2} \nabla_{\underline c} F_{\underline a \underline b}
\,\tilde{\bm{e}}^{\underline c} \wedge \tilde{\bm{e}}^{+} ~ , \nonumber\\
& \tilde{\bm{R}}_{\underline a +}
=-(\nabla_{\underline a} dV) \wedge \tilde{\bm{e}}^{+}+
\frac{q}{4} \nabla_{\underline a} F_{\underline b \underline c}
\,\tilde{\bm{e}}^{\underline b} \wedge \tilde{\bm{e}}^{\underline c}
-\frac{q^2}{4}F_{\underline b \underline c} F^{\underline b}{}_{\underline a} 
\,\tilde{\bm{e}}^{\underline c} \wedge \tilde{\bm{e}}^{+}~, \nonumber\\
& \tilde{\bm{R}}_{\underline a -}
= \tilde{\bm{R}}_{+-}=0~,
\end{align}
where $\bm{R}_{\underline a \underline b}$ are the curvature 2-forms on $(M,\bm{g})$.

The non-zero components of the Ricci curvature $\tilde{R}_{\underline A \underline B}$ are given by
\begin{equation}
\tilde{R}_{\underline a \underline b}=R_{\underline a \underline b}~,
\quad
\tilde{R}_{\underline a +}
=-\frac{q}{2} \nabla^{\underline b} F_{\underline b \underline a}~,
\quad
\tilde{R}_{++}=\frac{q^2}{4} F_{\underline a\underline b}
F^{\underline a\underline b}~.
\label{ricci_lift_basis}
\end{equation}

\subsection{Calculating the commutativity conditions}
\label{app:commutativity}

The commutativity conditions (\ref{commutativity})
are expressed by
\begin{equation}\label{anomaly}
\tilde{\delta} \tilde{\bm{m}}^{(i,j)}=0 \,,
\end{equation}
where
\begin{align}
\tilde{\bm{m}}^{(i,j)}
&= \frac{1}{2} \tilde{m}^{(i,j)}_{\underline {AB}} 
\,\tilde{\bm{e}}^{\underline A} \wedge \tilde{\bm{e}}^{\underline B} \nonumber\\
&= \frac{1}{2}
\tilde{m}^{(i,j)}_{\underline {ab}} \,\tilde{\bm{e}}^{\underline a} \wedge \tilde{\bm{e}}^{\underline b}
+\tilde{m}^{(i,j)}_{\underline a+} \,\tilde{\bm{e}}^{\underline a} \wedge \tilde{\bm{e}}^{+}
+\tilde{m}^{(i,j)}_{\underline a-} \tilde{\bm{e}}^{\underline a} \wedge \tilde{\bm{e}}^{-}
+\tilde{m}^{(i,j)}_{+-} \,\tilde{\bm{e}}^{+} \wedge \tilde{\bm{e}}^{-} \,.
\end{align}
Here, $\tilde{\delta}$ is the co-derivation on $(\tilde{M},\tilde{\bm{g}})$.
We note that, by use of the orthonormal basis (\ref{basis2_lift}) 
and the covariant derivatives (\ref{der1}), 
we may calculate the co-derivation $\tilde{\delta}$ as
\begin{equation}
\tilde{\delta}=- \tilde\eta^{\underline A \underline B} \tilde{\bm{e}}_{\underline A} 
\hook \tilde{\nabla}_{\tilde{\bm{e}}_{\underline B}} \,,
\end{equation}
where $\hook$ denotes the inner product.

Since we have ${\cal L}_{\tilde{\bm{e}}_\pm} \tilde{\bm{m}}^{(i,j)}=0$
and $\tilde{\bm{e}}^{\underline a}=\bm{e}^{\underline a}$,
it is natural to define the differential forms on $(M,\bm{g})$,
\begin{align}
\bm{m}^{(i,j)}
= \frac{1}{2} \tilde{m}^{(i,j)}_{\underline{ab}} 
\,\bm{e}^{\underline{a}} \wedge \bm{e}^{\underline b} \,, \qquad
\bm{U}^{(i,j)}_{+} = \tilde{m}^{(i,j)}_{\underline a +} \,\bm{e}^{\underline a} \,, \qquad
\bm{U}^{(i,j)}_{-} = \tilde{m}^{(i,j)}_{\underline a -} \,\bm{e}^{\underline a} \,,
\end{align}
and then we obtain
\begin{equation}
\tilde{\delta} \tilde{\bm{m}}^{(i,j)}
=
\delta \bm{m}^{(i,j)}
+
\left( \delta \bm{U}^{(i,j)}_{+}
-\delta^{\underline a \underline b}\tilde{m}^{(i,j)}_{\underline a-}
\nabla_{\underline b} V
+\frac{q}{2}
\tilde{m}^{(i,j)}_{\underline {ab}} F^{\underline {ab}} \right) \tilde{\bm{e}}^{+}
+\delta \bm{U}^{(i,j)}_{-} \,\tilde{\bm{e}}^{-}
\,,
\end{equation}
so that the commutativity conditions (\ref{anomaly}) are rewritten 
into the following equations:
\begin{align}
\delta \bm{m}^{(i,j)}=0 \,, 
\label{comcond1}\\
\delta \bm{U}^{(i,j)}_{+}
-\delta^{\underline a \underline b}\tilde{m}^{(i,j)}_{\underline a-}
\nabla_{\underline b} V
+\frac{q}{2}
\tilde{m}^{(i,j)}_{\underline {ab}} F^{\underline {ab}} =0 \,, 
\label{comcond2}\\
\delta \bm{U}^{(i,j)}_{-}=0 
\label{comcond3}\,.
\end{align}

In what follows, we calculate
the commutativity conditions (\ref{anomaly})
for the Killing tensors (\ref{K_uplifted_orthonormal}) on the uplifted spacetime.
To do so, we first calculate the 2-form $\tilde{\bm{m}}^{(i,j)}$.
After a straightforward calculation, we find that the components 
$\tilde{m}^{(i,j)}_{\underline\smu \hat{\underline\smu}}$,
$\tilde{m}^{(i,j)}_{\underline\smu+}$,
$\tilde{m}^{(i,j)}_{\underline a-}$,
and $\tilde{m}^{(i,j)}_{+-}$ are vanishing.
Hence, $\bm{U}_-^{(i,j)}$ vanishes and thus Eq.~(\ref{comcond3}) is satisfied.
Also, the second term of Eq.~(\ref{comcond2}) vanishes,
and $\bm{U}_+^{(i,j)}$
is given by
\begin{equation}
 \bm{U}^{(i,j)}_{+}
= \sum_{\mu=1}^n
\tilde{m}^{(i,j)}_{\hat{\underline \smu} +}
{\bm e}^{\hat{\underline \smu}}
+\epsilon \,
\tilde{m}^{(i,j)}_{\underline 0 +}
{\bm e}^{\underline 0}\,.
\end{equation}
Using the fact that the components
$\tilde{m}^{(i,j)}_{\hat{\underline \smu} +}$ 
and $\tilde{m}^{(i,j)}_{\underline 0 +}$
of $\bm{U}_+^{(i,j)}$ satisfy the conditions
\begin{equation}
 \bm{e}_{\hat{\underline \nu} }\tilde{m}^{(i,j)}_{\hat{\underline \smu} +}
 =\bm{e}_{\underline 0 }\tilde{m}^{(i,j)}_{\hat{\underline \smu} +}=0 \,, \qquad
 \bm{e}_{\hat{\underline \nu} }\tilde{m}^{(i,j)}_{\underline 0 +}
 =\bm{e}_{\underline 0 }\tilde{m}^{(i,j)}_{\underline 0 +}=0 \,,
\end{equation}
we can show that
\be
\delta \bm{U}_+^{(i,j)} = 0 \,.
\ee
Moreover, we can calculate that 
for the gauge field $\bm{A}$ by (\ref{A_LFKK}),
its field strength is given by
\begin{equation}
\bm{F}=d\bm{A}
=i\beta \frac{\partial \phi}{\partial x_\mu}
\,\bm{e}^{\underline\smu} \wedge \bm{e}^{\hat{\underline\smu}},
\end{equation}
where
\begin{equation}
\phi=\sum_{\rho=1}^n \frac{Q_\rho}{1+\beta^2 x_\rho^2}+\epsilon S .
\end{equation}
Here, $\epsilon=0$ for even dimensions and $\epsilon=1$ for odd dimensions.
While the nonzero components of $\bm{F}$ are $\underline{\mu\hat{\mu}}$ components only, 
all the $\underline{\mu\hat{\mu}}$ components 
of $\tilde{\bm{m}}^{(i,j)}$ vanish,
which leads to $\tilde{m}^{(i,j)}_{\underline{ab}}F^{\underline{ab}}=0$ and then Eq.~(\ref{comcond2}) is satisfied.
Finally, Eq.~(\ref{comcond1}) holds
since the Killing tensors
on the Kerr-NUT-(A)dS spacetime commute with each other
(see, e.g., \cite{Kolar:2015cha}).

\bibliography{Eisenhartlift_2018}
\bibliographystyle{JHEP}

\end{document}